\documentclass[aps,prb,twocolumn,floatfix, superscriptaddress]{revtex4-1}
\usepackage{times}
\usepackage{epsfig}
\usepackage{amsmath}
\usepackage{amssymb}
\usepackage{graphicx}
\usepackage{dsfont}
\usepackage[utf8]{inputenc}
\usepackage[colorlinks=true,citecolor=red,linkcolor=blue]{hyperref}

\begin{document}
\title{Quench dynamics in the one-dimensional mass-imbalanced ionic Hubbard model}
\author{Zhuotao Xie}
\affiliation{College of Physics and Technology, Guangxi Normal University, Guilin, Guangxi 541004, China}
\author{Ming Zhao}
\affiliation{College of Physics and Technology, Guangxi Normal University, Guilin, Guangxi 541004, China}
\author{Hantao Lu}
\affiliation{School of Physical Science and Technology \& Lanzhou Center for Theoretical Physics, Key Laboratory of Theoretical Physics of Gansu Province, Lanzhou University, Lanzhou 730000, China}
\author{Zhongbing Huang}
\affiliation{Department of Physics, Hubei University, Wuhan 430062, China}
\author{Gregory A. Fiete}
\affiliation{Department of Physics, Northeastern University, Boston, Massachusetts 02115, USA}
\affiliation{Department of Physics, Massachusetts Institute of Technology, Cambridge, Massachusetts 02139, USA}
\author{Xiang Hu}
\affiliation{College of Physics and Technology, Guangxi Normal University, Guilin, Guangxi 541004, China}
\author{Liang Du}
\email{liangdu@gxnu.edu.cn}
\affiliation{College of Physics and Technology, Guangxi Normal University, Guilin, Guangxi 541004, China}

\begin{abstract}
Using the time-dependent Lanczos method, we study the non-equilibrium dynamics of the one-dimensional ionic-mass imbalanced Hubbard chain driven by a quantum quench of the on-site Coulomb interaction, where
the system is prepared in the ground state of the Hamiltonian with a different Hubbard interaction.
A full exact diagonalization is adopted to study the zero temperature phase diagram in equilibrium, which is shown to be in good agreement with previous studies using density matrix renormalization group (DMRG).
We then study the non-equilibrium quench dynamics of the spin and charge order parameters by fixing the initial and final Coulomb interaction while changing the quenching time protocols.  The Lanczos method allows us to reach longer times following the quench than DMRG.
Our study shows that the time evolution of the charge and spin order parameters strongly depend on the quenching time protocols. In particular, the effective temperature of the system will decrease monotonically as the quenching time is increased.
Finally, by taking the final Coulomb interaction strength to be in the strong coupling regime, we find that the oscillation frequency of the charge order parameter increases monotonically with the Coulomb interaction. By contrast, the frequency of the spin order parameter decreases monotonically with increasing Coulomb interaction. We explain this result using an effective spin model in the strong coupling limit. Our study suggests strategies to engineer the relaxation behavior of interacting quantum many-particle systems.
\end{abstract}
\date{\today}
\maketitle

\section{Introduction}
The understanding of non-equilibrium dynamics of a strongly correlated electronic system has seen dramatic progress from both theoretical and experimental sides in the past decade\cite{Kollar:pra2008,Rigol:nat2008,Aoki:rmp2014,Eckardt:rmp2017,Torre:rmp2021}.
Two commonly studied scenarios are the laser driven strongly correlated solid state system
and a Coulomb interaction quenched system in optical lattices with cold atomic gases\cite{Aoki:rmp2014,Mentink:nc2015,Bukov:prl2016, Eckardt:rmp2017,Puviani:prb2016,Ono:prl2017,Claassen:nc2017, LiuJ:prl2018,Kennes:prl2018,Seetharam:prb2018,Messer:prl2018, Sandholzer:prl2019,Kaneko:prl2019,Vogl:prx2019,Kiffner:prb2019,Chaudhary:prb2019,Oka:arcmp2019,LiJ:prl2020, GaoH:prl2020, Torre:rmp2021,Quito:prl2021,Arakawa:prb2021, Kumar:prb2021}. In driven systems, the observation of hidden quantum states not accessible in equilibrium\cite{Stojchevska:science2014}, and the non-equilibrium control of quantum phase transitions in correlated electron systems, have attracted great interest. For example, the ac-field drive dynamical band flipping\cite{Tsuji:prl2011}, the damping of Bloch oscillations in the Falicov-Kimball model\cite{Freericks:prl2006,Freericks:prb2008} and Hubbard model\cite{Eckstein:prl2011}, the ultra-fast control of magnetic order in the Mott insulators\cite{Mentink:nc2015,Ono:prl2017,Chaudhary:prb2019}, and
photo-induced unconventional superconductivity\cite{Fausti:science2011,Kitamura:prb2016,Kaneko:prl2019} illustrate known phenomena.

In these non-equilibrium systems, the study of long-time thermalization behavior is of particular interest.
In general, a closed (driven) system will thermalize to a featureless infinite temperature thermal state with maximal entropy if the energy of the system is not conserved\cite{Alessio:prx2014,Lazarides:prl2014, Lazarides:pre2014}, unless the system is sufficiently disordered for many-body localization\cite{Ponte:prl2015,Lazarides:prl2015}.
If the system is coupled to a bath (i.e., ``open"), it is possible to establish a non-equilibrium steady state, since the absorbed energy can be released to the connected bath\cite{Seetharam:prx2015,Shirai:pre2015}.
For a clean isolated solid state system driven by spatially uniform electric field, the system could show different thermalization behavior, resulting in a featureless infinite temperature steady state, a non-thermal steady state, or even an oscillatory state\cite{Fotso:scirep2014,Joura:prb2015,Fotso:fp2020}.
In the case of periodic driving, the heating rate can depend on the laser frequency.
Abanin {\it et al.}\cite{Abanin:prl2015,Abanin:prb2017} find the heating rate decreases exponentially as the driving frequency is increased, provided the frequency is larger than other characteristic energy scales in the Hamiltonian.
Mallayya {\it et al.}\cite{Mallayya:prx2019,Mallayya:prl2019,Mallayya:prb2021} confirm the robust exponential regime using a numerical linked-cluster expansion method,
and suggest the heating rate should obey Fermi's golden rule in a weakly perturbed non-integrable system.
Seetharam {\it et al.}\cite{Seetharam:prb2018} find that the Floquet eigenstates in a clean system can exhibit non-thermal behavior because of a finite system size.

In general, a Coulomb interaction quenched system will thermalize unless the system is integrable\cite{Rigol:prl2007,Cazalilla:prl2006,Kollar:pra2008,Iucci:pra2009,Du:prb2018a}. In a quench from a superfluid to a Mott regime, the system will thermalize in some regimes while not in others\cite{Rigol:pra2009}.
A numerical study of a finite quantum system of bosons found that the thermalization behavior depends on the magnitude of Coulomb interaction change\cite{Roux:pra2009} and the distance in parameter space to an integrable point, with a failure to thermalize as one approaches the integrable point\cite{Rigol:pra2009}
By contrast, the quenched fermionic Hubbard model on an infinite dimensional Bethe-lattice system will result in a quasi-stationary state for a weak and strong Coulomb interaction quench, while in between the two regimes, a dynamical phase transition is observed, where fast thermalization occurs\cite{Eckstein:prl2009,Eckstein:prb2010a,Du:prb2017c}.

In this work, we are interested in the one-dimensional ionic mass imbalanced fermionic Hubbard model, which has been studied using mean-field theory (MFT)\cite{Sekania:prb2017,Yen:cp2019} and the density matrix renormalization group (DMRG)\cite{Padilla-Gonzalez:arxiv2021} method.
Compared to the conventional Hubbard model, the translational and spin SU(2) symmetry are explicitly broken, while the mass imbalance breaks the SU(2) symmetry, the ionic term (staggered potential) breaks the translational symmetry.
There exist two phases in the plane of the Coulomb interaction and the on-site ionic term for fixed mass imbalance: (i) a charge density wave order induced by the staggered ionic potential and (ii) an alternating magnetic order originating from the hopping asymmetry and Coulomb interaction. The transition at finite $U_c$ is characterized as first order at the mean-field level\cite{Sekania:prb2017} and second order in DMRG\cite{Padilla-Gonzalez:arxiv2021}. If one extends the above one-dimensional model to a two-dimensional square lattice, one finds novel magnetically ordered metallic phases in the Coulomb interaction and the staggered potential plane\cite{Chattopadhyay:prb2021}. For example, a spin imbalanced ferromagnetic metal, a ferromagnetic metal, and an anti-ferromagnetic half metal all appear. However, the non-equilibrium behavior of this model has not been studied until now.

In this paper, we study the non-equilibrium dynamics in the one dimensional ionic Hubbard model while quenching the Coulomb interaction, where the initial state is prepared in the ground state of an initial Hamiltonian.
From the technical point of view, DMRG works exceptionally well in the static case, while time-dependent DMRG suffers from significant growth of the entanglement entropy when studying quench dynamics\cite{Kloss:prb2018,Goto:prb2019}.
In our work, we adopt the time-dependent Lanczos method in studying quench dynamics of the  ionic mass imbalanced Hubbard model in one-dimension.

The experimental realization of the ionic mass imbalanced Hubbard model can be implemented in ultra-cold atoms in engineered optical lattice systems.
The hopping asymmetry (mass imbalance) can be introduced by considering two species of fermionic atoms (e.g., $^6$Li and $^{40}$K) trapped in an optical lattice\cite{Cazalilla:prl2005}, where the staggered ionic potential can be created by the interference of counter-propagating laser beams\cite{Chattopadhyay:prb2021}, and the Coulomb interaction strength can be tuned via a magnetic Feshbach resonance\cite{Wille:prl2008,Esslinger:arcmp2010,Buchler:prl2010}.

\begin{figure*}[ht]
\centering
\includegraphics[angle=-90,width=0.325\textwidth]{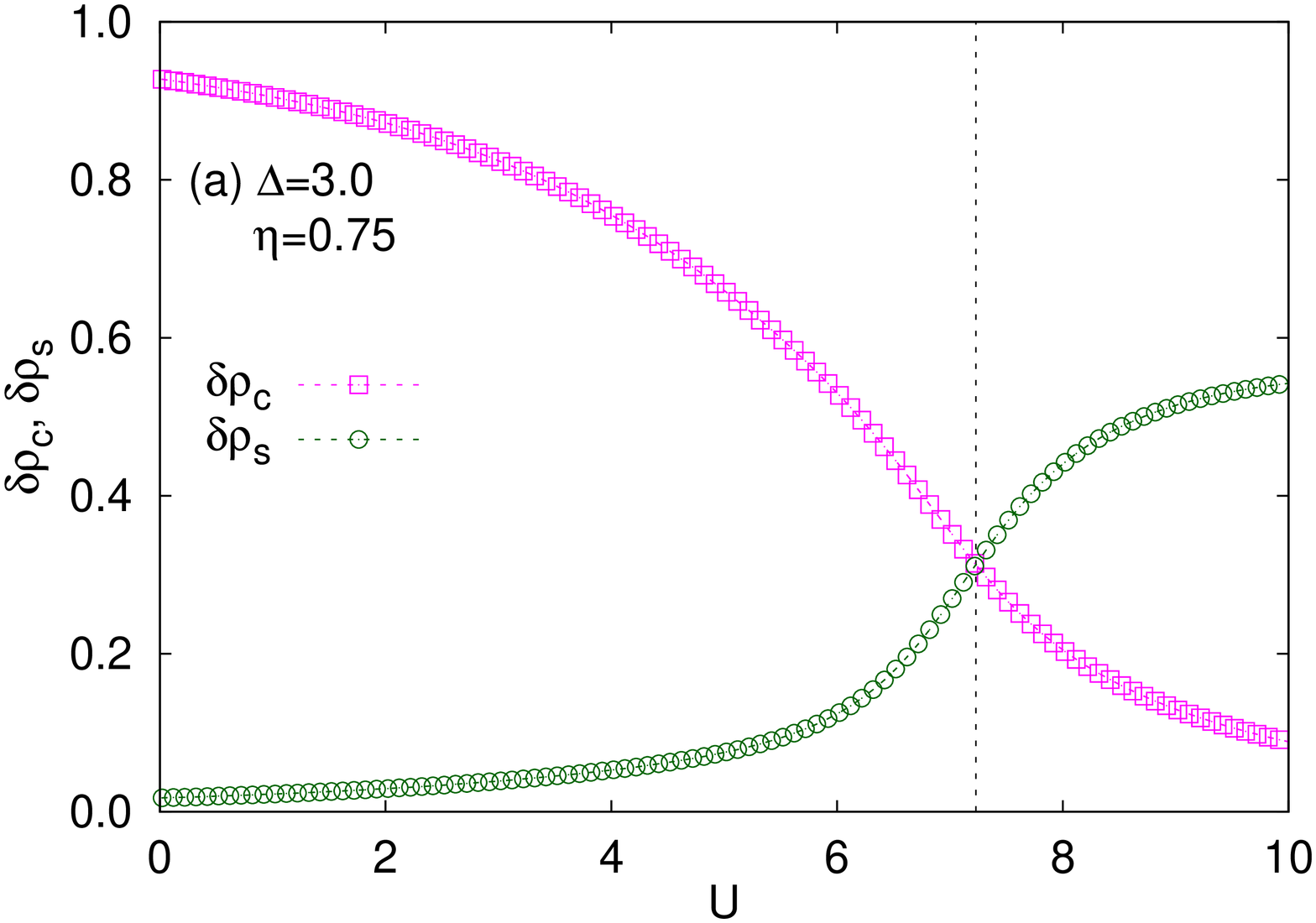}
\includegraphics[angle=-90,width=0.325\textwidth]{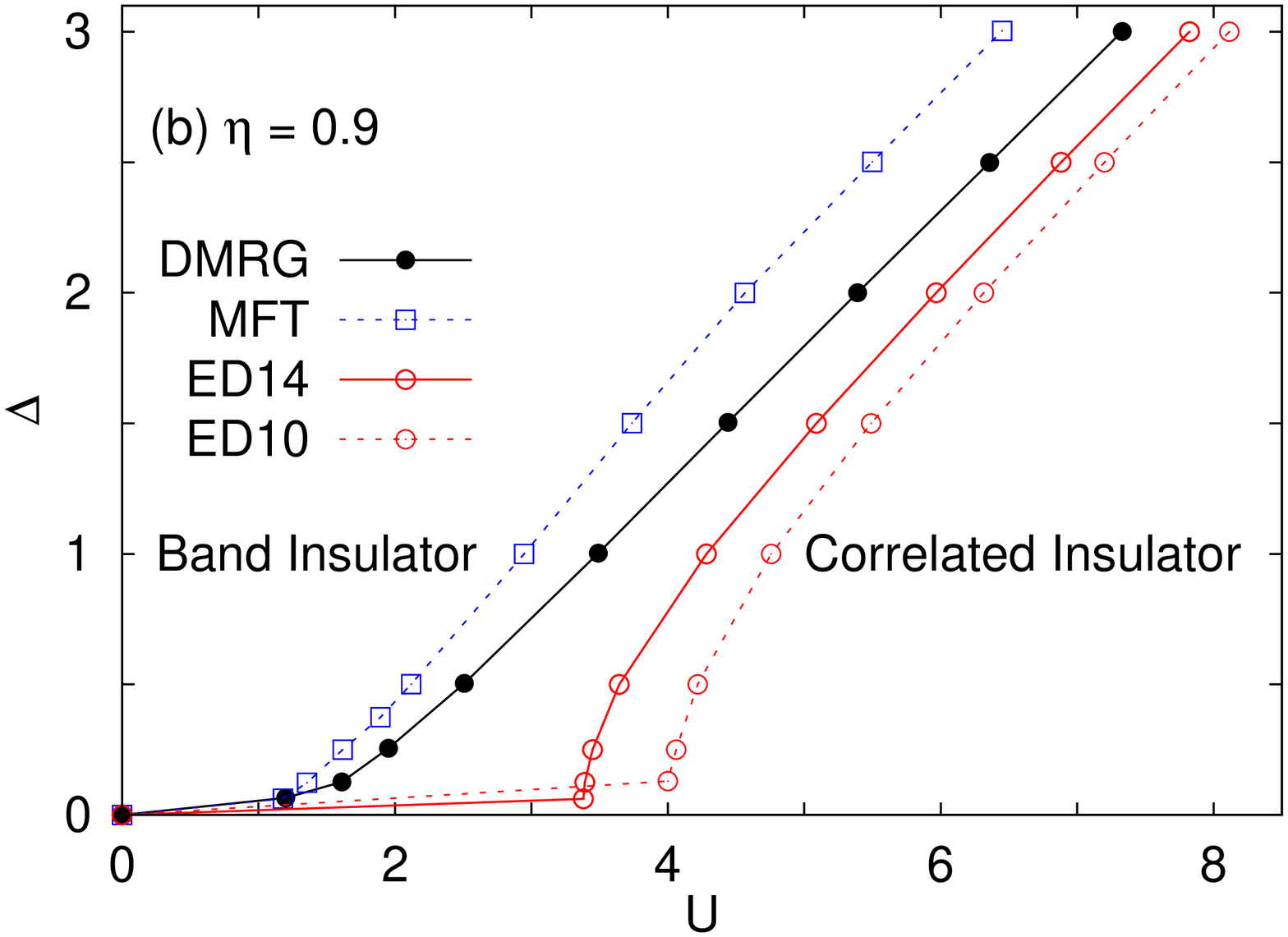}
\includegraphics[angle=-90,width=0.325\textwidth]{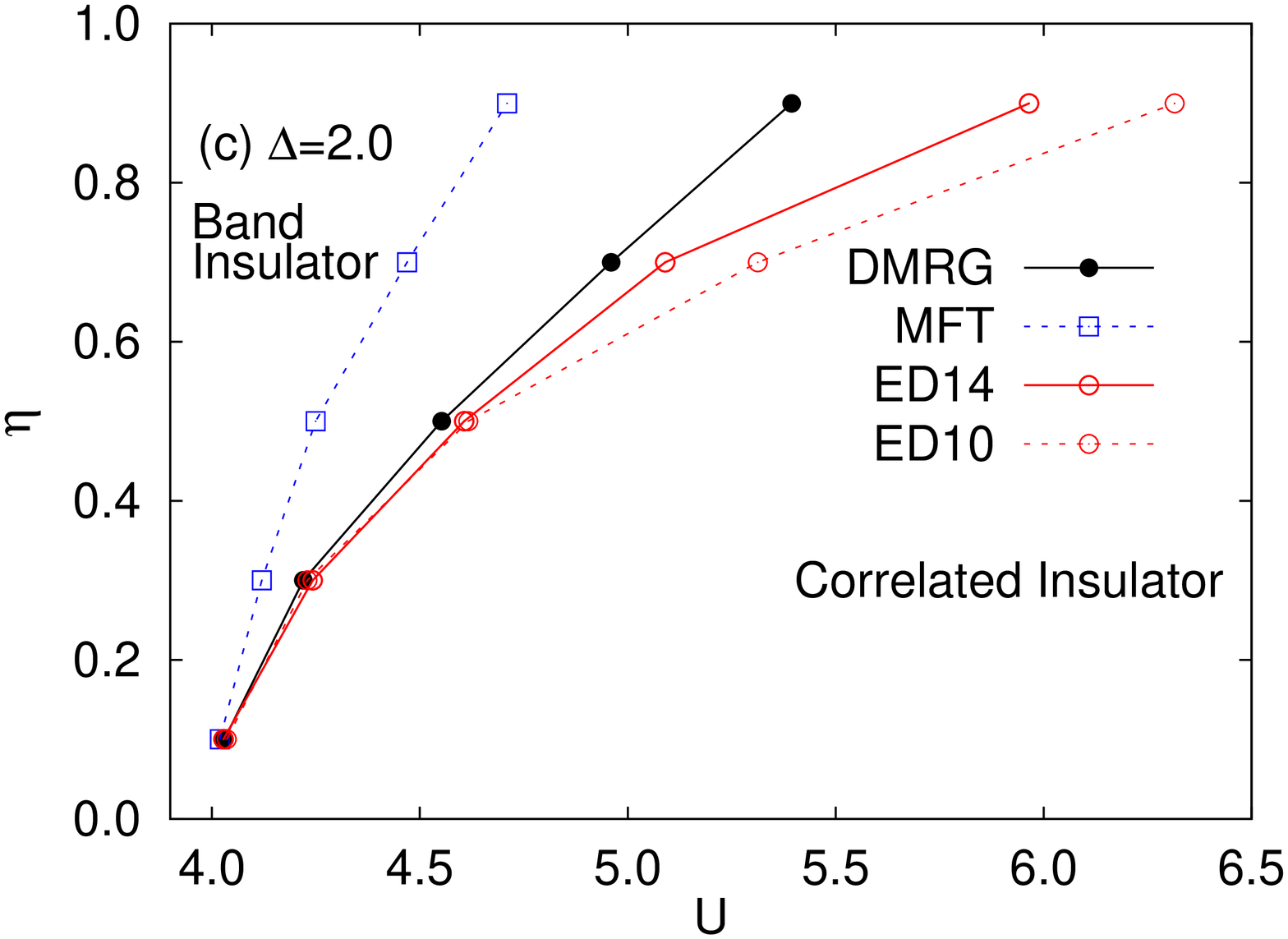}
\caption{(Color online)
The equilibrium phase diagram of the one-dimensional mass imbalanced Hubbard chain at half-filling and zero temperature,  calculated using exact diagonalization (10 sites and 14 sites).
The density matrix renormalization group (DMRG) and mean-field theory (MFT) data are obtained from the references [\onlinecite{Sekania:prb2017, Padilla-Gonzalez:arxiv2021}]. (a) The order parameters $\delta\rho_s$ and $\delta\rho_c$ as a function Coulomb interaction are plotted to characterize the phase transition (dashed line at $U_c=7.1$), where the mass imbalance is $\eta = 0.75$, and crystal field $\Delta=3.0$.
(b) Critical points are plotted in the plane of Coulomb interaction and crystal field with fixed mass imbalance $\eta = t_{\downarrow}/t_{\uparrow} = 0.9$.
(c) Critical points are plotted in the plane of Coulomb interaction and mass imbalance $\eta$ with fixed crystal field $\Delta = 2.0$.
}
\label{fig1}
\end{figure*}

Our paper is organized as follows.
In Sec.\ref{model}, we describe the Hamiltonian of the one-dimensional mass imbalanced ionic Hubbard model and the time-dependent Lanczos method.
In Sec.\ref{edphase}, the equilibrium phase diagram is obtained using exact diagonalization.
In Sec.\ref{qdyn}, we calculate the non-equilibrium quench dynamics of the system in different Coulomb interaction regimes.
Finally, in Sec.\ref{concl} we present the main conclusions of the paper.

\section{Model Hamiltonian and time-dependent Lanczos algorithm}
\label{model}
The time dependent mass imbalanced ionic Hubbard model in one dimension is,
\begin{align}
     H(t) = &-\sum_{i\sigma}t_h^{\sigma}\left(c_{i,\sigma}^{\dagger}c_{i+1,\sigma}^{}
             +c_{i+1,\sigma}^{\dagger}c_{i,\sigma}^{}\right)\nonumber\\
         &+\Delta\sum_{i\sigma}(-1)^i n_{i\sigma}+U(t)\sum_{i}n_{i\uparrow}n_{i\downarrow},
\label{Htime}
\end{align}
where $c_{i\sigma}^{\dagger}$ ($c_{i\sigma}$) creates (annihilate) an electron with (pseudo) spin $\sigma$ at site $i$ ($i=1,\cdots, L$), and $n_{i\sigma}=c_{i\sigma}^{\dagger}c_{i\sigma}$ is the corresponding occupancy operator.   Here, $t_h^\sigma$ is the hopping integral between nearest-neighbors for spin $\sigma$ electron, $-\Delta$ ($\Delta$) is the ionic potential for odd (even) sites of the one dimensional chain, and
$U(t)$ is the time dependent on-site Coulomb interaction.

Throughout this paper, we set $t_h^\uparrow = 1$ as the unit of energy and the time is in units of $1/t_h^\uparrow$, correspondingly.
The hopping asymmetry (mass imbalance) is defined as the ratio of spin-$\downarrow$ to spin-$\uparrow$ hopping integrals $\eta = t_h^{\downarrow} / t_h^{\uparrow}$.
In the following, we restrict ourselves to the half-filling case with periodic boundary conditions,
where the total number of electrons $N$ is equal to the number of sites in the chain $L$.
Furthermore, we assume the total magnetization in the system vanishes,
which means the number of up spin electrons $N_{\uparrow}$ is equal to the down spin electrons $N_{\downarrow}$.
The non-equilibrium quench dynamics is studied by fixing the hopping parameter $t_h^{\downarrow}/t_h^{\uparrow}\leq 1$ and ionic potential $\Delta \geq 0$
while quenching the Coulomb interaction from an initial $U(t=0^-) = U_i$ to final $U(t \geq t_q) = U_f$, where $t_q$ is the linear ramp time of Coulomb interaction change.

The exact diagonalization method (a standard Lanczos procedure) is employed to numerically find the ground state of the Hamiltonian at time $t=0^-$ where $U(t=0^-) = U_i$.
This state is used as an initial state for the time dependent Schr\"odinger equation $i\partial_t |\Psi(t)\rangle = H(t) |\Psi(t)\rangle$.
The time evolution is implemented step-by-step based on the time-dependent Lanczos method\cite{Park:jcp1986,Mohankumar:cpc2006,Balzer:jpcm2012,Lu:prl2012,Innerberger:epjp2020},
\begin{equation}
  |\Psi(t+\delta t)\rangle \approx e^{-i H(t)\delta t} |\Psi(t)\rangle
                           \approx \sum_{l}^M e^{-i \epsilon_l \delta t} |\Phi_l\rangle\langle \Phi_l| \Psi(t)\rangle,\nonumber
\end{equation}
where $\epsilon_l$ ($\Phi_l$) are the eigenvalues (eigenvectors) of the tri-diagonal matrix generated by Lanczos iteration with $M \leq 100$. (In general $M=30$ works well.)
We set $\delta t = 0.005$ in our calculation of the time evolution.
The physical observable is computed as,
\begin{align}
     \langle O(t)\rangle = \langle \Psi(t)| O |\Psi(t)\rangle.
\end{align}
Following earlier work\cite{Sekania:prb2017}, we measured the evolution of the charge and spin density order parameters,
\begin{align}
     \delta\rho_{c}(t) &=-\frac{1}{L}\sum_{i\sigma}(-1)^i \langle \Psi(t)| n_{i\sigma} |\Psi(t)\rangle, \nonumber\\
     \delta\rho_{s}(t) &=\frac{1}{L}\sum_{i\sigma}\sigma(-1)^i \langle \Psi(t)| n_{i\sigma}|\Psi(t)\rangle,
\end{align}
where $\sigma = 1 (-1)$ for spin-$\uparrow$ ($\downarrow$) electrons in the second line.
The effective temperature after Coulomb interaction quench protocol is calculated by numerically solving the equation\cite{Eckstein:prb2010a},
\begin{equation}
E(t \geq t_q) = \frac{\mathrm{Tr}[H(t\geq t_q) e^{-H(t\geq t_q)\beta_{\mathrm{eff}}}]}{\mathrm{Tr}[e^{-H(t\geq t_q)\beta_{\mathrm{eff}}}] }
\label{Teff}
\end{equation}
where $E(t \geq t_q)$ is the energy after the Coulomb interaction quench protocol and the effective temperature is denoted as $T_{\mathrm{eff}} = 1/ \beta_{\mathrm{eff}}$. To solve Eq.\eqref{Teff} numerically, a finite temperature Lanczos algorithm\cite{Jaklic:prb1994,LiJ:cpb2022} is used to calculate the total energy of the equilibrium system at finite temperature.

\section{Equilibrium Phase Diagram at Zero Temperature}
\label{edphase}
To check the validity of exact diagonalization (ED), we calculate the equilibrium phase diagram and compare it with the data obtained with DMRG or MFT.
In Fig.\ref{fig1}(a), we plot the charge and spin order parameter as a function of Coulomb interaction for the one-dimensional mass imbalanced Hubbard chain with 14 sites and periodic boundary conditions.
The mass imbalance is set as $\eta = t_{h}^\downarrow/t_{h}^\uparrow = 0.75$.

In the non-interacting limit, the charge and spin order parameter is derived analytically as $\delta\rho_c = 0.927$ and $\delta\rho_s = 0.018$, respectively.
With increasing Coulomb interaction, the charge order parameter $\delta\rho_c$ will decrease while the spin order parameter $\delta\rho_s$ increases monotonically.
At the critical Coulomb interaction $U_c = 7.1$, we have $\delta\rho_c = \delta\rho_s$.
Further increasing the Coulomb interaction will result in $\delta\rho_c < \delta\rho_s$, which is the correlated insulator phase\cite{Sekania:prb2017,Padilla-Gonzalez:arxiv2021}.

This behavior above can be understood from the two limits of the Coulomb interaction strength.
In the non-interacting limit, the model can be solved analytically and the two order parameters are expressed as a function of an elliptic integral\cite{Sekania:prb2017}, where  the charge order parameter increases monotonically as a function of crystal field and finally converges to $\delta\rho_c=1$.
Conversely, the spin order parameter will decrease monotonically and converge to 0.
In the strong Coulomb interaction limit $U \gg t_\uparrow, t_\downarrow, \Delta$, the system will be reduced to an anisotropic XXZ Heisenberg model with a staggered magnetic field,
which result in an anti-ferromagnetic Mott insulating phase with $\delta\rho_s \approx 1$ and $\delta\rho_c \approx 0$.

The effective Hamiltonian in the strong coupling limit is\cite{Grusha:ijmpb2016},
\begin{align}
    H_{\mathrm{eff}} &= \mathcal{J}_\mathrm{ex}\sum_i\left(S_i^x S_{i+1}^x + S_i^y S_{i+1}^y + \gamma S_i^z S_{i+1}^z\right)  \nonumber\\
    &\quad - h\sum_{i} (-1)^i S_i^z,
\end{align}
where $S_i^{x(y,z)}$ is the spin operator at the $i$-th site along  direction $x(y,z)$ and the coupling coefficients are,
\begin{small}
\begin{align}
    \mathcal{J}_\mathrm{ex} = \frac{4U t_{\uparrow}t_{\downarrow}}{U^2 - 4\Delta^2}, \quad
    \gamma = \frac{t_\uparrow^2 + t_{\downarrow}^2}{2t_{\uparrow}t_{\downarrow}},\quad
    h = \frac{4(t_\uparrow^2 - t_{\downarrow}^2)\Delta}{U^2 - 4\Delta^2}.
    \label{Jexdef}
\end{align}
\end{small}
Here, $\gamma \neq 1$ breaks the SU(2) symmetry and $h\neq 0$ breaks the translational symmetry.
In Fig.\ref{fig1}(b), the phase diagram in the plane of crystal field $\Delta$ and Coulomb interaction $U$ is plotted
with fixed mass imbalance $\eta = t_{\downarrow}/t_{\uparrow} = 0.9$.
The critical points are characterized by the crossing of the two order parameters\cite{Sekania:prb2017} and can be confirmed by studying the von Neumann block entropy as a function of the Coulomb interaction\cite{Padilla-Gonzalez:arxiv2021}.
The band insulator and correlated insulator phase are observed in the large crystal field and large Coulomb interaction regime, respectively.
To validate the exact diagonalization method used in this paper, we plot the phase diagram derived by the Hartree-Fock mean-field method and DMRG method as a comparison.

Compared to the data obtained with DMRG, exact diagonalization overestimates the critical Coulomb interaction for a fixed crystal field $\Delta$, which we attribute to a finite size effect:
Increasing the chain size from $L=10$ to 14 decreases the deviation.
Furthermore, our numerical calculations show that the difference decreases with increasing $\Delta$.
To check the dependence on the mass imbalance, the phase diagram in the plane of mass imbalance $\eta$ and Coulomb interaction $U$ for a fixed crystal field $\Delta = 2.0$ is plotted in Fig.\ref{fig1}(c).
Comparing to the phase diagram calculated by DMRG, one can see that ED has worked well for large mass imbalance.

\section{Non-equilibrium quench dynamics}
\label{qdyn}
In our study of non-equilibrium quench dynamics, we fix the mass imbalance $\eta = t_\downarrow/t_\uparrow = 0.75$ and crystal field $\Delta = 3.0$ while changing the Coulomb interaction strength $U(t)$. The quench protocol is defined through the time dependent Coulomb interaction as,
\begin{equation}
    U(t) = \begin{cases}
          U_i + \alpha t & t < t_q \\
          U_f, & t \geq  t_q \\
          \end{cases},
\end{equation}
where $\alpha = (U_f - U_i)/t_q$ is the slope of ramp in the Coulomb quench protocol.
The starting state is set as the ground state of the initial equilibrium Hamiltonian with $U(t=0^-)=U_i$. The time evolution of the initial state is based on the time-dependent Hamiltonian in Eq.\eqref{Htime}.
\begin{figure*}[ht]
\centering
\includegraphics[angle=-90,width=0.325\textwidth]{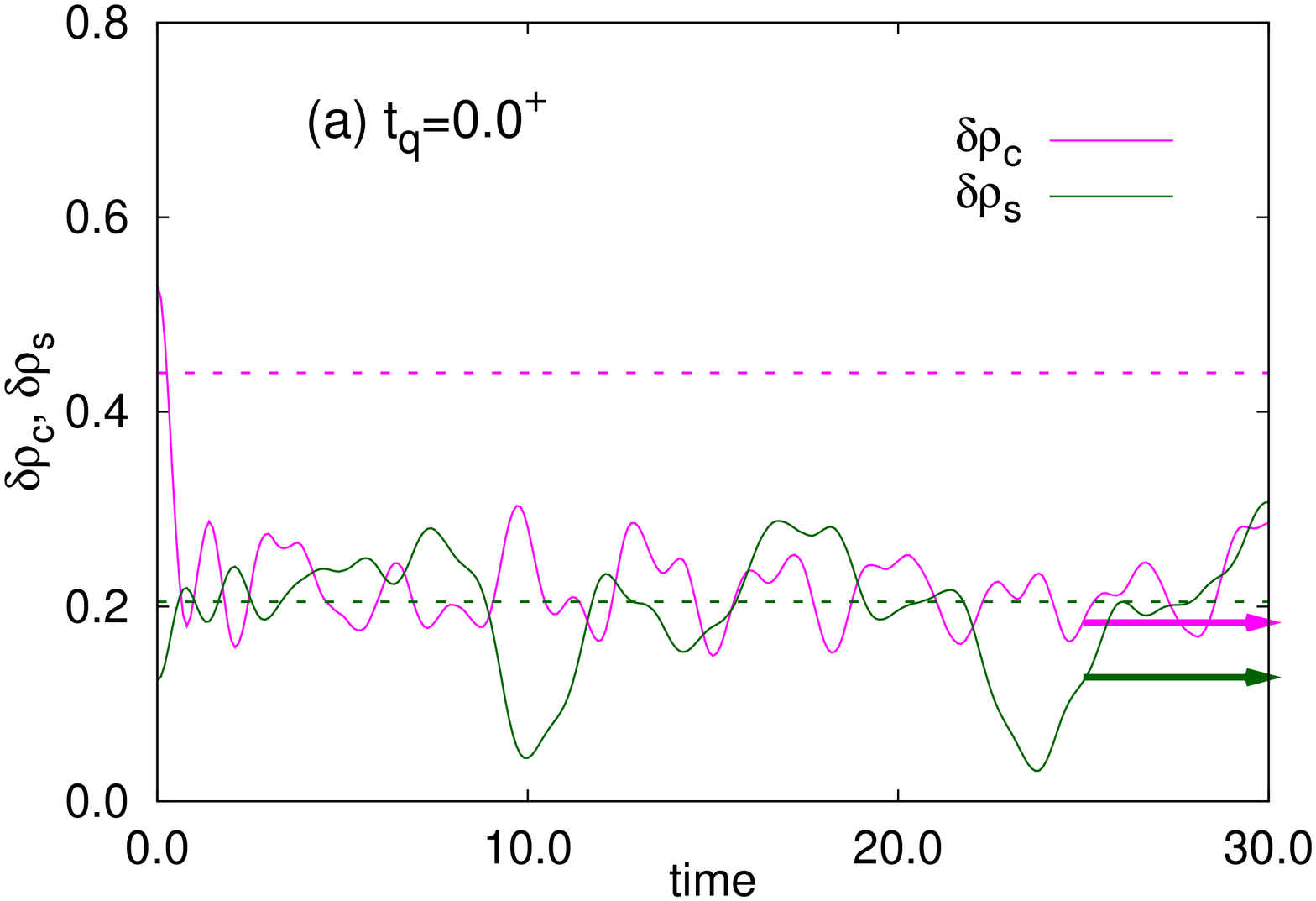}
\includegraphics[angle=-90,width=0.325\textwidth]{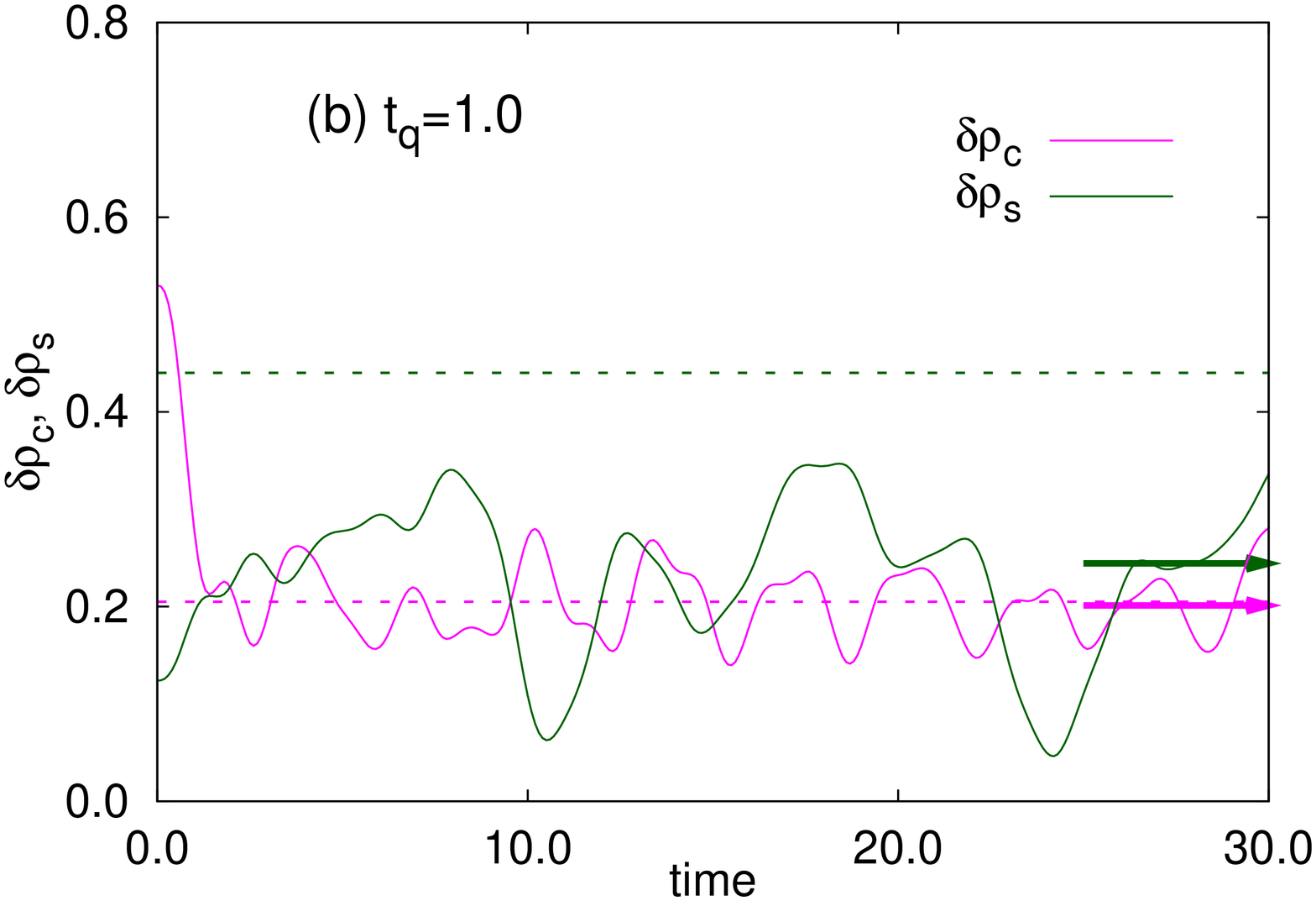}
\includegraphics[angle=-90,width=0.325\textwidth]{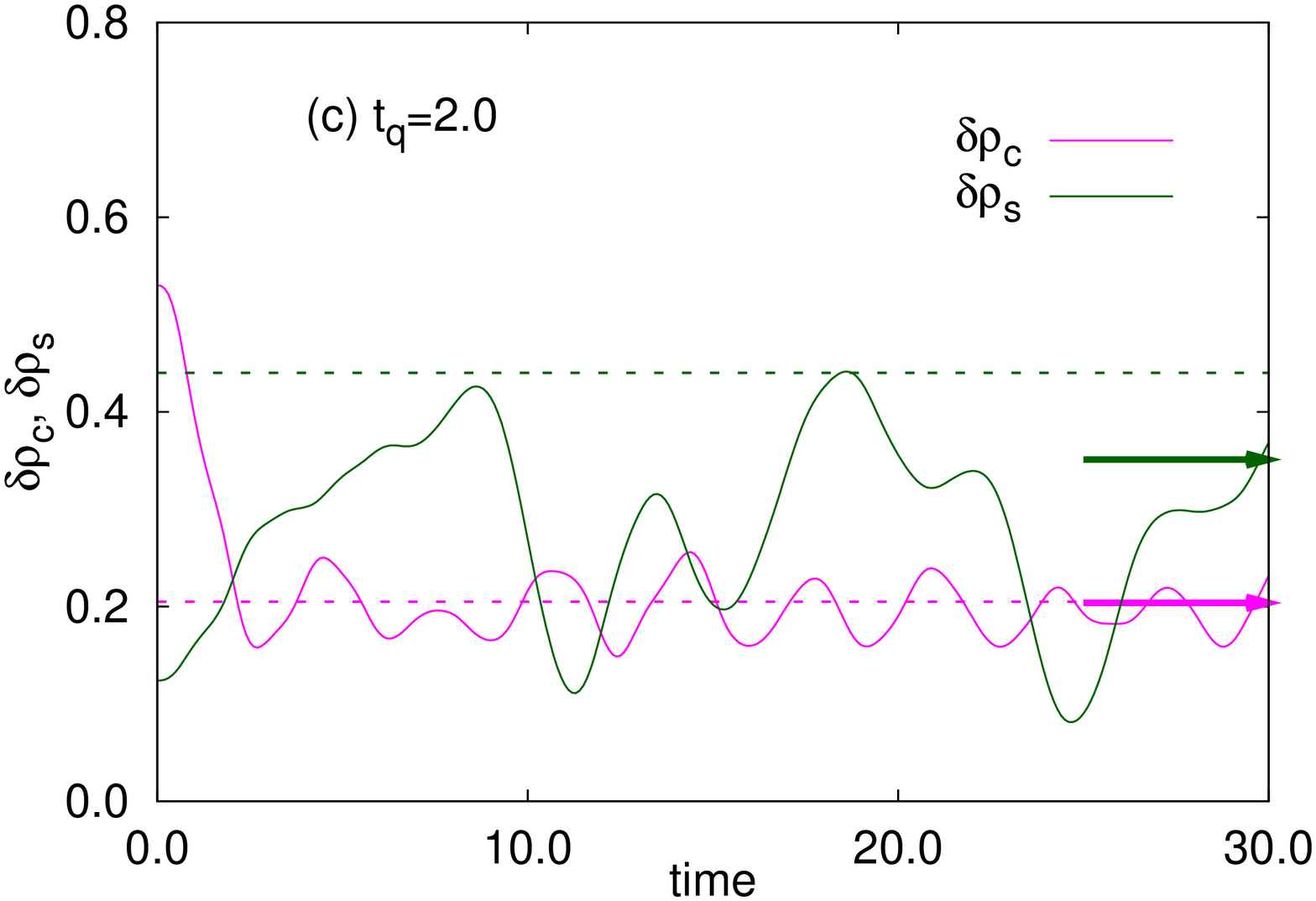}
\includegraphics[angle=-90,width=0.325\textwidth]{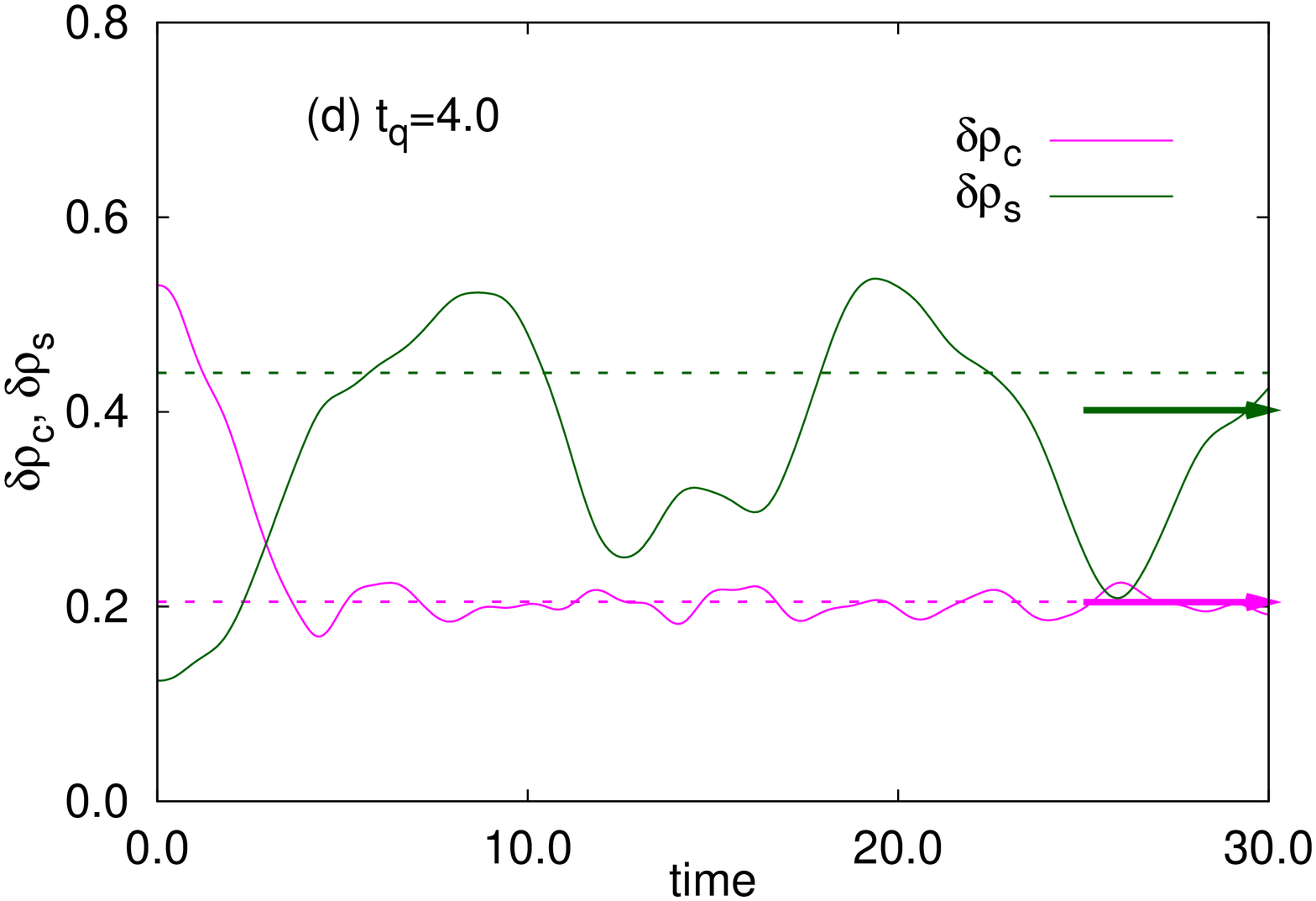}
\includegraphics[angle=-90,width=0.325\textwidth]{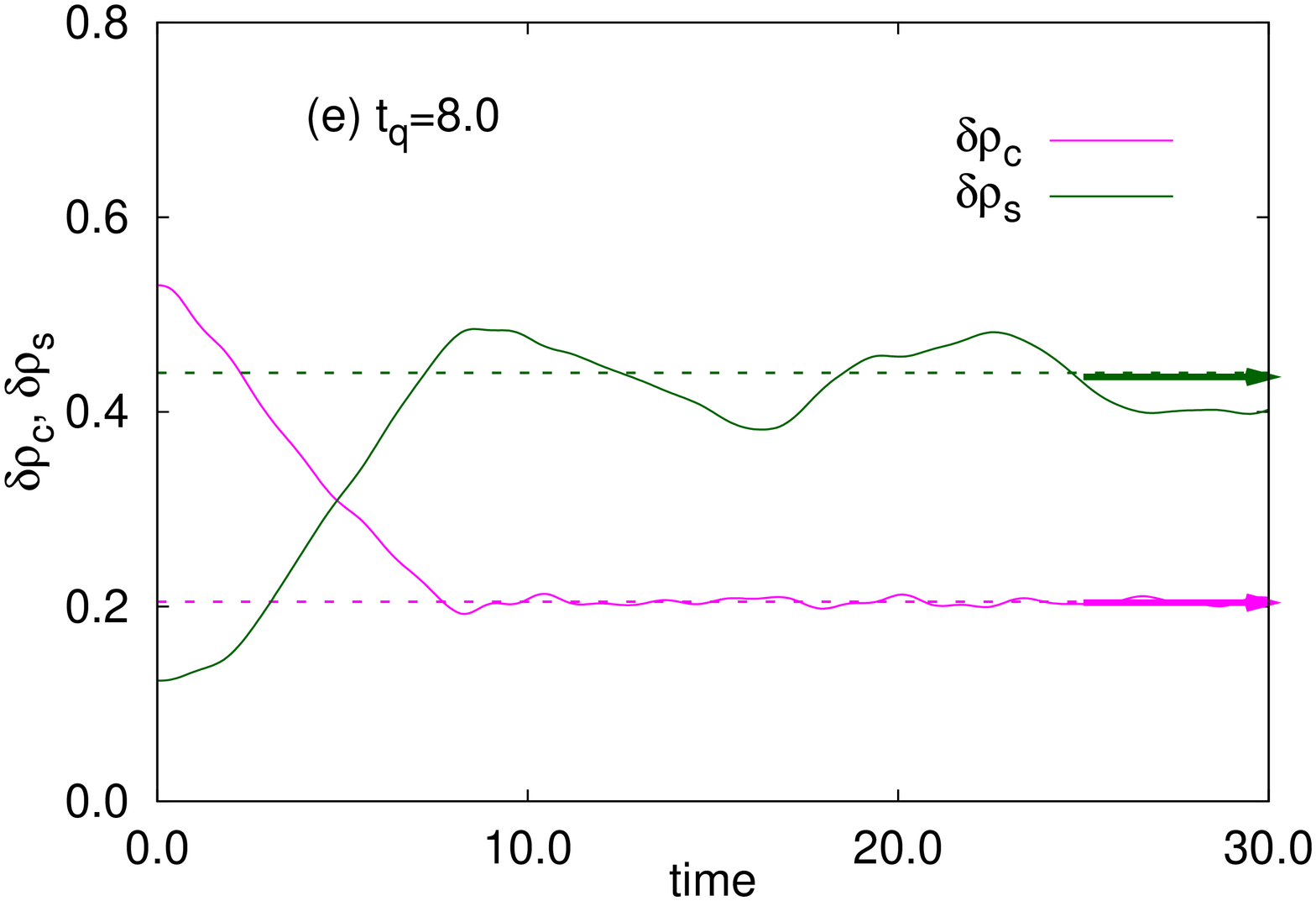}
\includegraphics[angle=-90,width=0.325\textwidth]{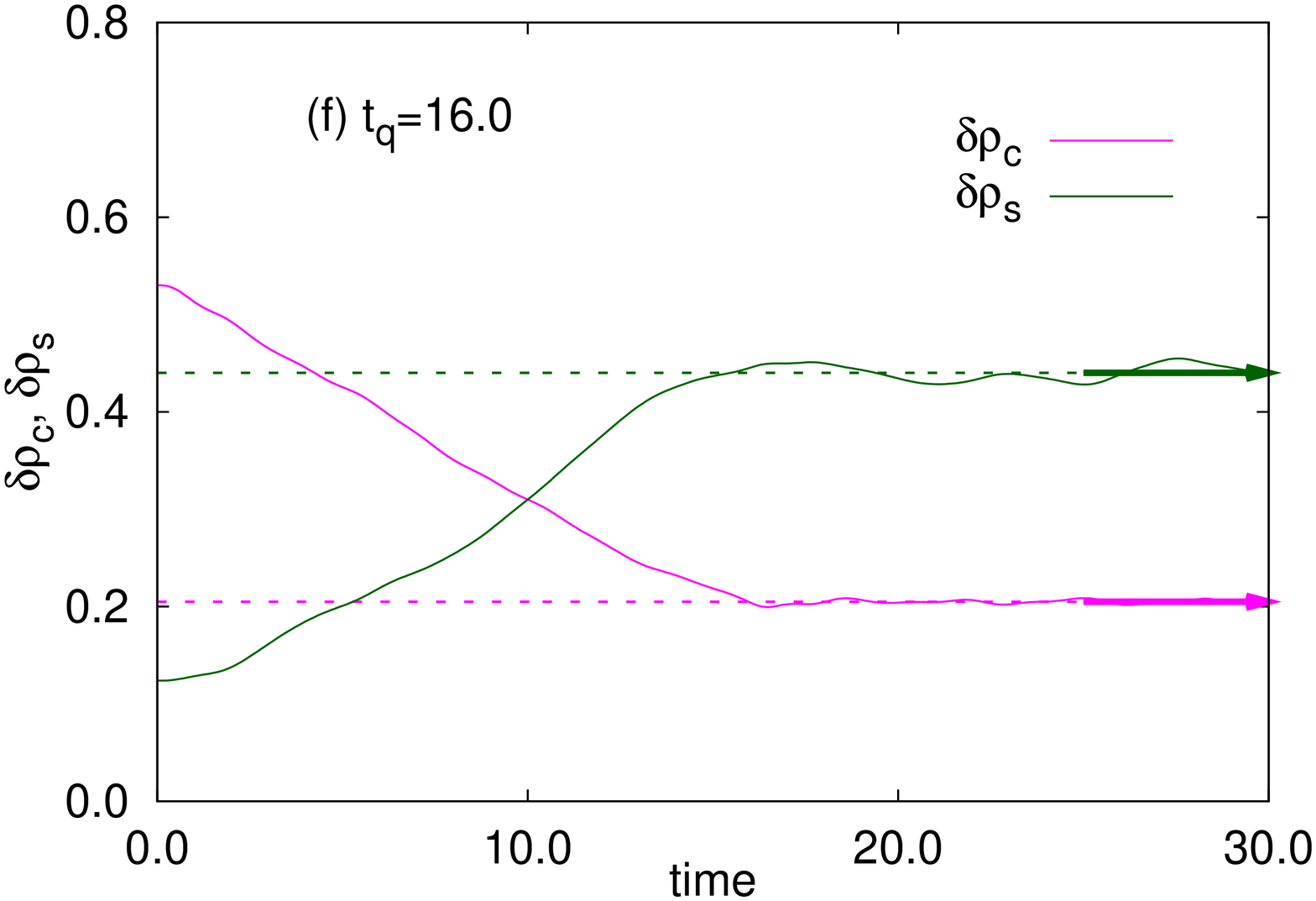}
\caption{(color online)
Time evolution of the charge and spin order parameter $\delta\rho_c$ and $\delta\rho_s$ after the Coulomb interaction quench protocol $U(t) = U_i + (U_f - U_i)t/t_q$ for $t < t_q$ and $U(t) = U_f$ for $t \geq t_q$, where the initial and final Coulomb interaction strength is set as $U_i=6.0$ and $U_f=8.0$, respectively.
(a) $t_q = 0.0^+$, (b) $t_q = 1.0$, (c) $t_q = 2.0$, (d) $t_q = 4.0$, (e) $t_q = 8.0$, (f) $t_q = 16.0$.
The expectation value of $\delta\rho_c$ and $\delta\rho_s$ at zero temperature is represented by a dashed line with Coulomb interaction $U = U_f$ in the equilibrium calculation. The arrows indicate the thermal values of the two order parameters at effective temperature $T_{\mathrm{eff}} = 0.628,\; 0.433,\; 0.270,\; 0.179,\; 0.105,\; 0.008$ for $t_q=0.0^+, \;1.0,\; \cdots,\; 16.0$, respectively.
}
\label{fig2}
\end{figure*}

\subsection{Non-equilibrium quench dynamics - Dependence on ramp time in the quench protocol}
To study the dependence of the ramp time in the quench protocol, we fix the initial and final Coulomb interaction $U_i, U_f$ while changing the quench time $t_q$.
In our calculation, we set the initial and final Coulomb interaction as $U_i = 6.0$ and $U_f = 8.0$, where the critical interaction between band and correlated insulator is $U_c = 7.1$.
In Fig.\ref{fig2}, we plot the non-equilibrium evolution of the charge and spin order parameters in the Hamiltonian, Eq.\eqref{Htime}, with different quench time $t_q=0.0^+,\; 1.0,\; 2.0,\; 4.0,\; 8.0,\; 16.0$.
The zero temperature ground state expectation values of $\delta\rho_c$ and $\delta\rho_s$ in equilibrium with Coulomb interaction $U = U_f$ are represented with dashed line $\delta\rho_c = 0.205$ and $\delta\rho_s = 0.440$.
The initial order parameters in equilibrium ($t=0$) are $\delta\rho_c^i = 0.530$ and $\delta\rho_s^i = 0.124$, respectively.

In the case with quench time $t_q = 0.0^+$ in Fig.\ref{fig2}(a) (the quench protocol is defined via a Heaviside step function),
the two order parameters are intertwined with each other and the phase transition between band insulator and correlated insulator can not be distinguished clearly.
The charge order parameter as a function of time oscillates around $\delta\rho_c = 0.199$, which is close to the equilibrium value at zero temperature, $\delta\rho_c^{eq} = 0.205$.
By contrast, the spin order parameter oscillates about $0.10$, which is far from the equilibrium value at zero temperature $\delta\rho_s^{eq} = 0.440$.
By using the definition of effective temperature Eq. \eqref{Teff}, we find $T_{\mathrm{eff}} = 0.476$. If the equilibrium calculation is done with $U=U_f$ at temperature $T = 0.476$, we have $\delta\rho_c^{th} = 0.181$ and $\delta\rho_s^{th} = 0.107$ which is represented as arrows in the plot.
By defining the oscillating amplitude as the difference between the maximum and the minimum of the order parameter after $t > 20$,
we find the amplitude of charge order parameter is about  $A(\delta\rho_c) = 0.092$, which is smaller than the spin order parameter $A(\delta\rho_s)= 0.241$.

In Fig.\ref{fig2}(b), we plot the time evolution of the two order parameters with a different quench time, $t_q = 1.0$.
In the quenching time regime $t < t_q$, a monotonically increasing  behavior of spin order and monotonically decreasing charge order parameter is observed. Compared to the case with $t_q = 0.0^+$, the oscillation amplitude of the spin and charge order parameter are larger with $A(\delta\rho_c) =0.134$ and $A(\delta\rho_s) \approx 0.283$. The effective temperature is $T_{\mathrm{eff}} = 0.432$. The order parameters at the effective temperature are $\delta\rho_c^{th} = 0.196$ and $\delta\rho_s^{th} = 0.204$, and are indicated by the arrows on the right side of the figure.

In Fig.\ref{fig2}(c-f), the quench protocol is changed by increasing the quenching time as $t_q = 2.0,\;4.0,\; 8.0,\; 16.0$.
Upon further increasing the quench time, the oscillation amplitude will decrease and the oscillation center moves closer to the one with zero temperature in equilibrium.  The calculated effective temperatures are $T_{\mathrm{eff}} =0.270,\; 0.179,\; 0.105,\; 0.050$, from which we conclude that the effective temperature decreases with increasing quench time, $t_q$.

The systematic quench time behavior can be understood as longer quench times $t_q$ making the Hubbard $U$ increase closer to adiabatic evolution, and therefore inducing less heating.  By checking the difference of the thermal values (arrows) and the expectation values at zero temperature (dashed line), we find that the spin order parameter $\delta\rho_s$ is much more sensitive to the effective temperature (greater for smaller $t_q$) than the charge order parameter $\delta\rho_c$.

\begin{figure*}[ht]
\centering
\includegraphics[angle=-90,width=0.325\textwidth]{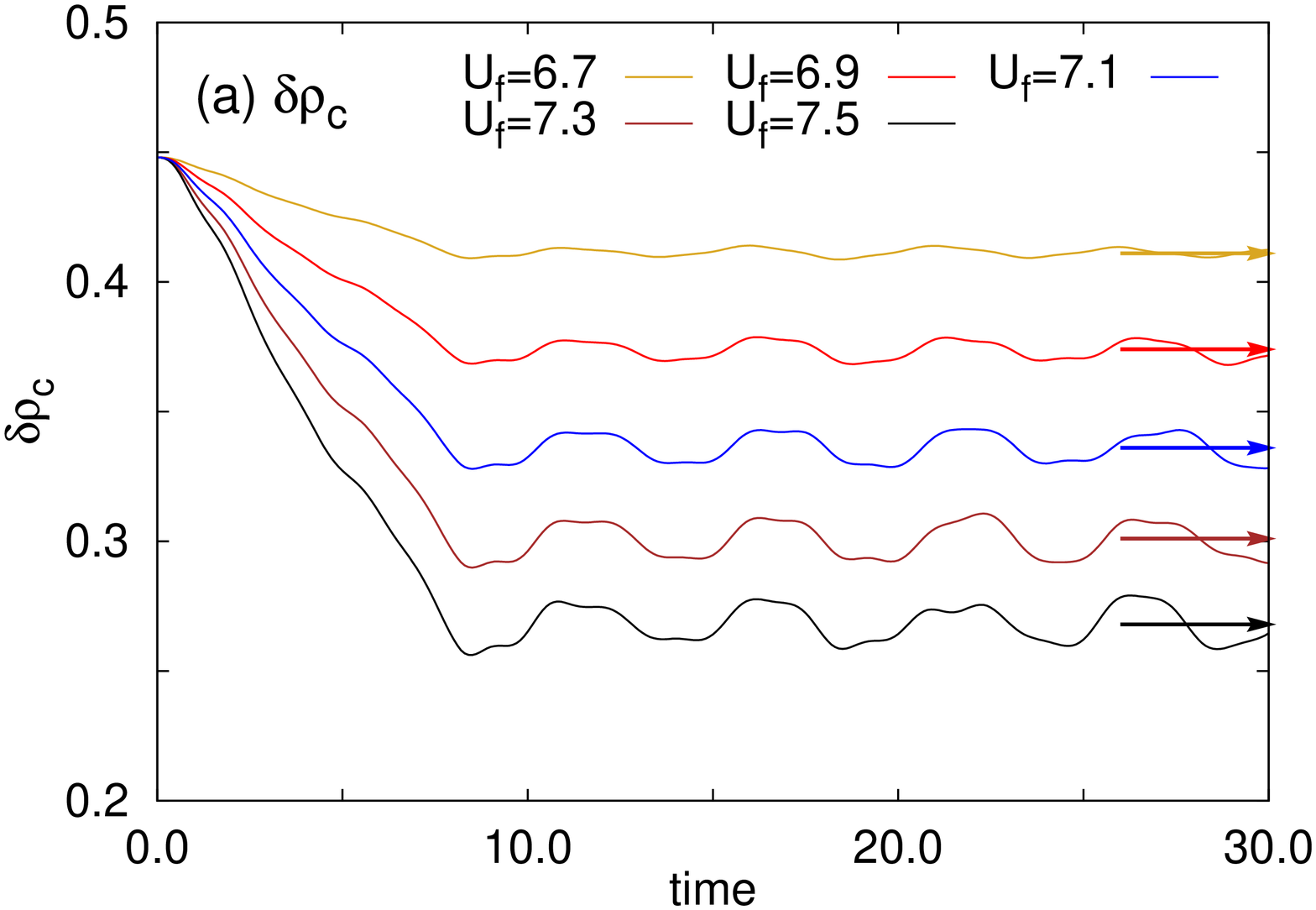}
\includegraphics[angle=-90,width=0.325\textwidth]{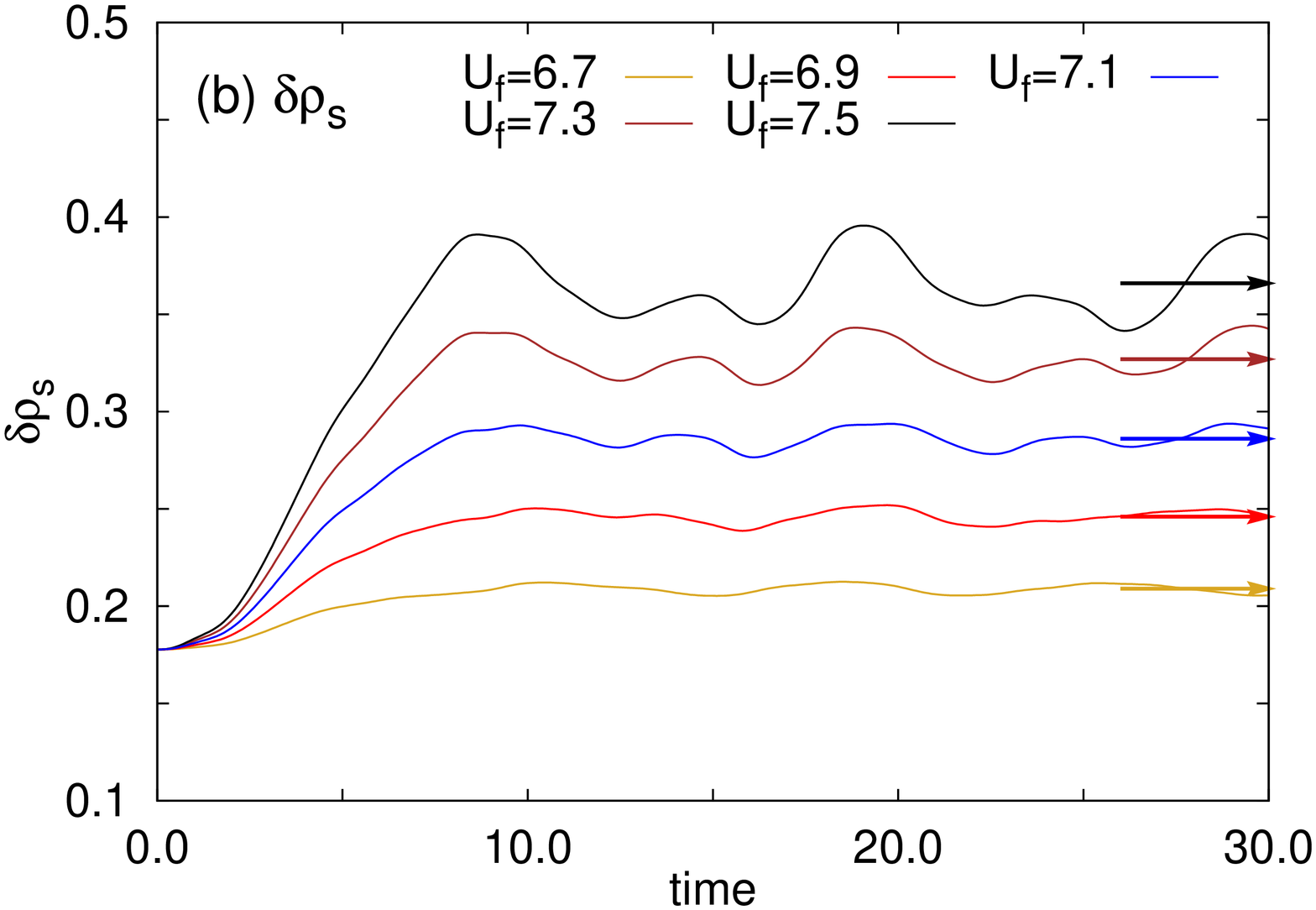}
\includegraphics[angle=-90,width=0.325\textwidth]{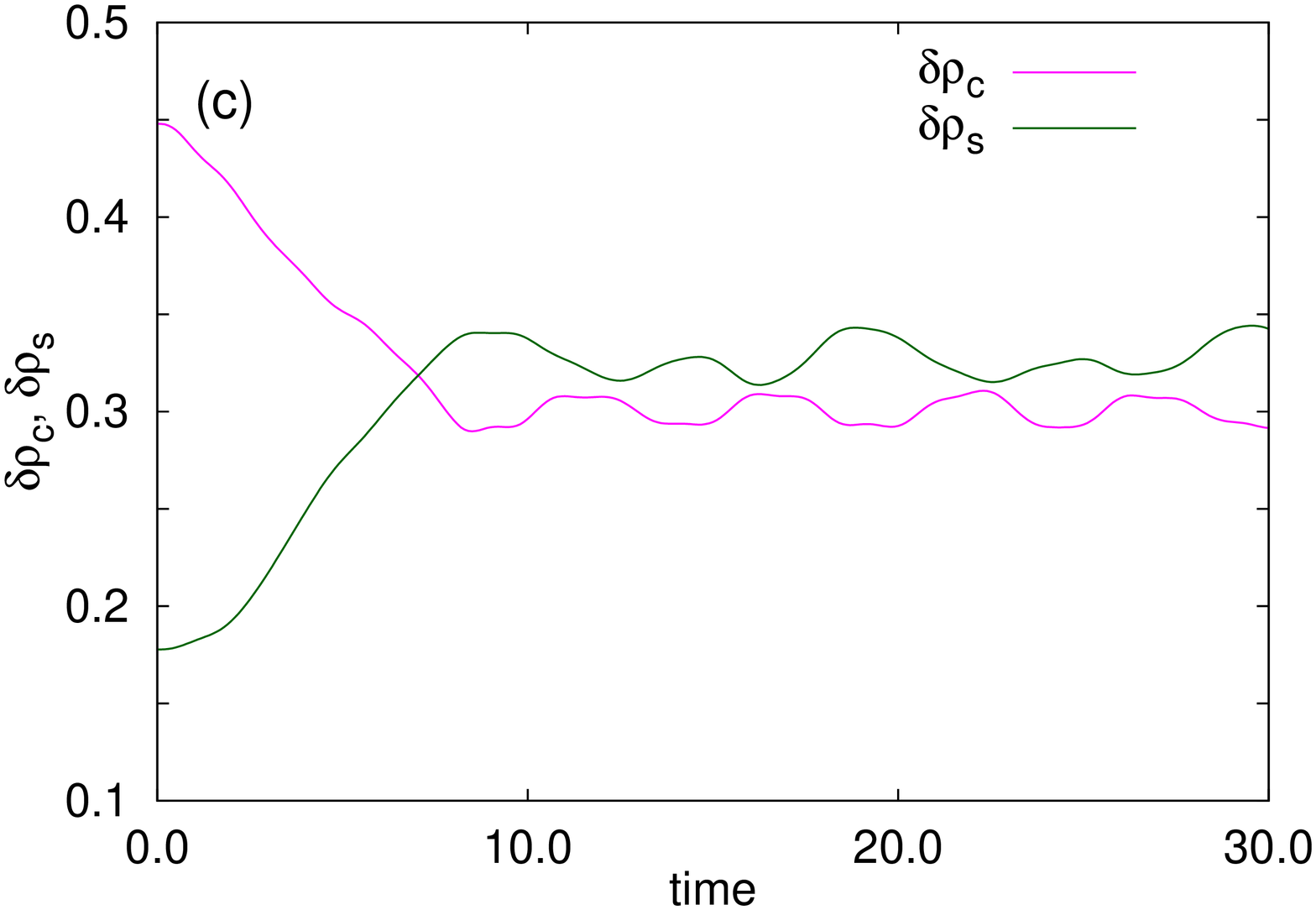}
\caption{(Color online) Time evolution of charge and spin order parameter $\delta\rho_c$ (a) and $\delta\rho_s$ (b) for Coulomb quenches from $U_i=6.5$ to $U_f=6.7,6.9,\cdots,7.5$, (c) Comparison of  $\delta\rho_c$ and $\delta\rho_s$ for quench from $U_i=6.5$ to $U_f=7.3$.
}
\label{fig3}
\end{figure*}

\subsection{Non-equilibrium quench dynamics near the critical Coulomb interaction}
For the case with mass imbalance $\eta = t_{\downarrow}/t_{\uparrow} = 0.75$ and crystal field $\Delta = 3.0$, the critical Coulomb interaction is $U_c = 7.1$ for the transition from band insulator to correlated insulator in equilibrium. To investigate the quench dynamics around the critical Coulomb interaction, we set the initial Coulomb interaction as $U_i = 6.5$ and the final Coulomb interaction to be near the critical value, $U_c = 7.1$. The quenching time is fixed at $t_q = 8.0$ while the slope for the quench protocol is $(U_f - U_i)/t_q$ is different for each specific $U_f$.

In Fig.\ref{fig3}(a-b), we plot the time evolution of the charge and spin order parameters after a Coulomb interaction quench from $U_i=6.5$ to different $U_f = 6.7,\; 6.9,\; 7.1,\; 7.3,\; 7.5$. The charge and spin order parameters are $\delta\rho_c = 0.448$ and $\delta\rho_s = 0.178$ for the initially prepared equilibrium system with $U_i = 6.5$. In the time range $0< t < t_q$, the charge (spin) order parameter decreases (increases) monotonically, which can be attributed to the increase of the Coulomb interaction. Over the time range $t > t_q$, the two order parameters oscillate with different amplitudes.  The amplitudes are summarized in Table.I, which are 0.005,  0.010, 0.013, 0.017, 0.021 for charge order parameters and 0.007, 0.010, 0.014, 0.031, and 0.058 for spin order parameters.

To compare the two order parameters, we plot the two in Fig.\ref{fig3}(c) with a final Coulomb interaction of $U_f = 7.3$. The only crossing of the two order parameters is observed at time $t=7.06$, which indicates a stable phase transition.
\begin{table}[htbp]
  \begin{center}
  \renewcommand\arraystretch{1.2}
  \setlength{\tabcolsep}{4mm}
    \caption{Oscillation amplitude of the charge and spin order parameters for quench protocols with different final $U_f$.  The ramp time $t_q$ and initial interaction value $U_i$ are kept fixed.}
    \begin{tabular}{ccccc} 
      \hline
      \hline
      $t_q$ &$U_i$ & $U_f$ & $A(\delta\rho_c)$ & $A(\delta\rho_s)$ \\
      \hline
      8.0 & 6.5 & 6.7 & 0.005 & 0.007\\
      \hline
      8.0 & 6.5 & 6.9 & 0.010 & 0.010\\
      \hline
      8.0 & 6.5 & 7.1 & 0.013 & 0.014\\
      \hline
      8.0 & 6.5 & 7.3 & 0.017 & 0.031\\
      \hline
      8.0 & 6.5 & 7.5 & 0.021 & 0.058\\
      \hline
    \end{tabular}
  \end{center}
  \label{table1}
\end{table}
\begin{figure*}[ht]
\centering
\includegraphics[angle=-90,width=0.235\textwidth]{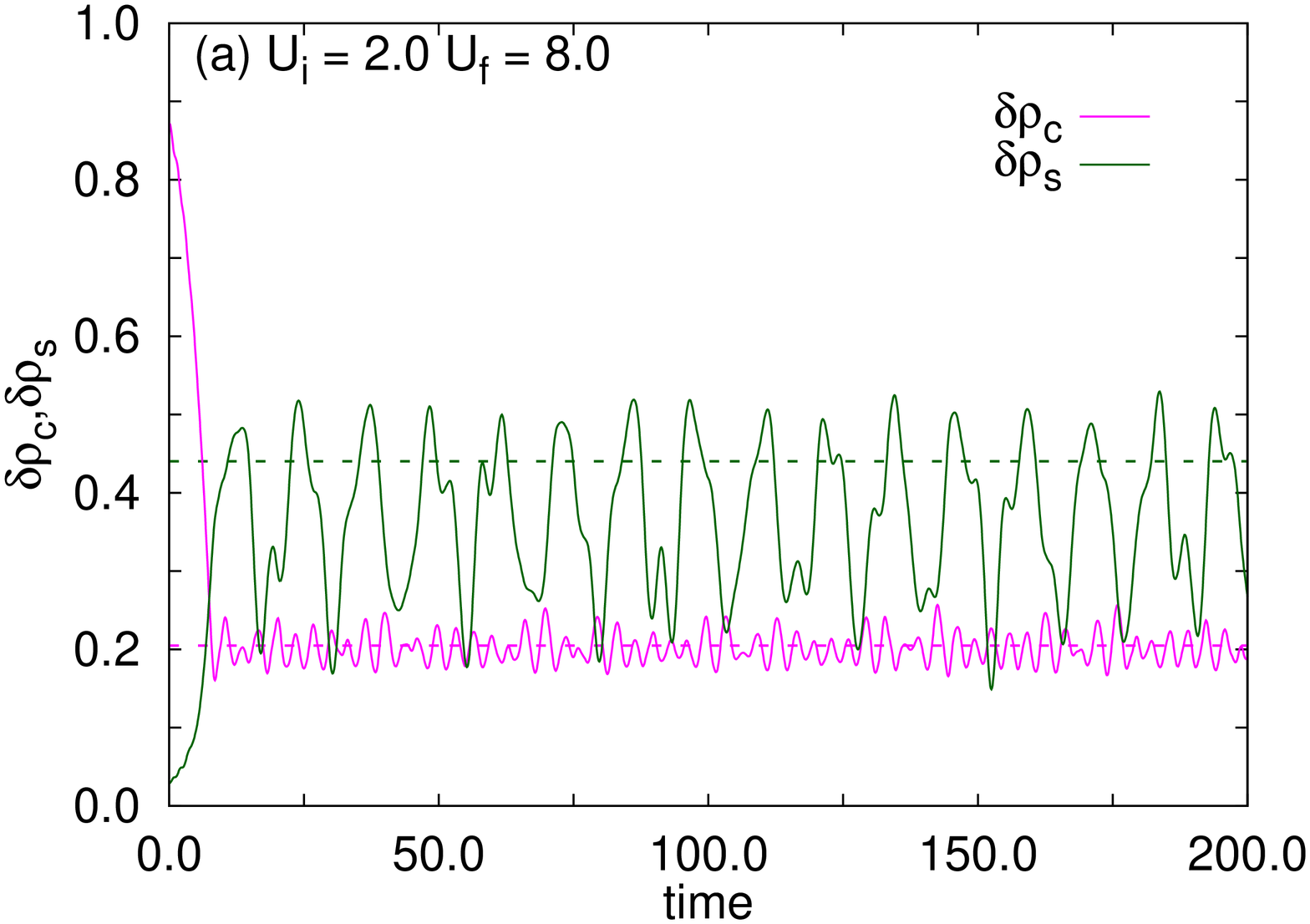}
\includegraphics[angle=-90,width=0.235\textwidth]{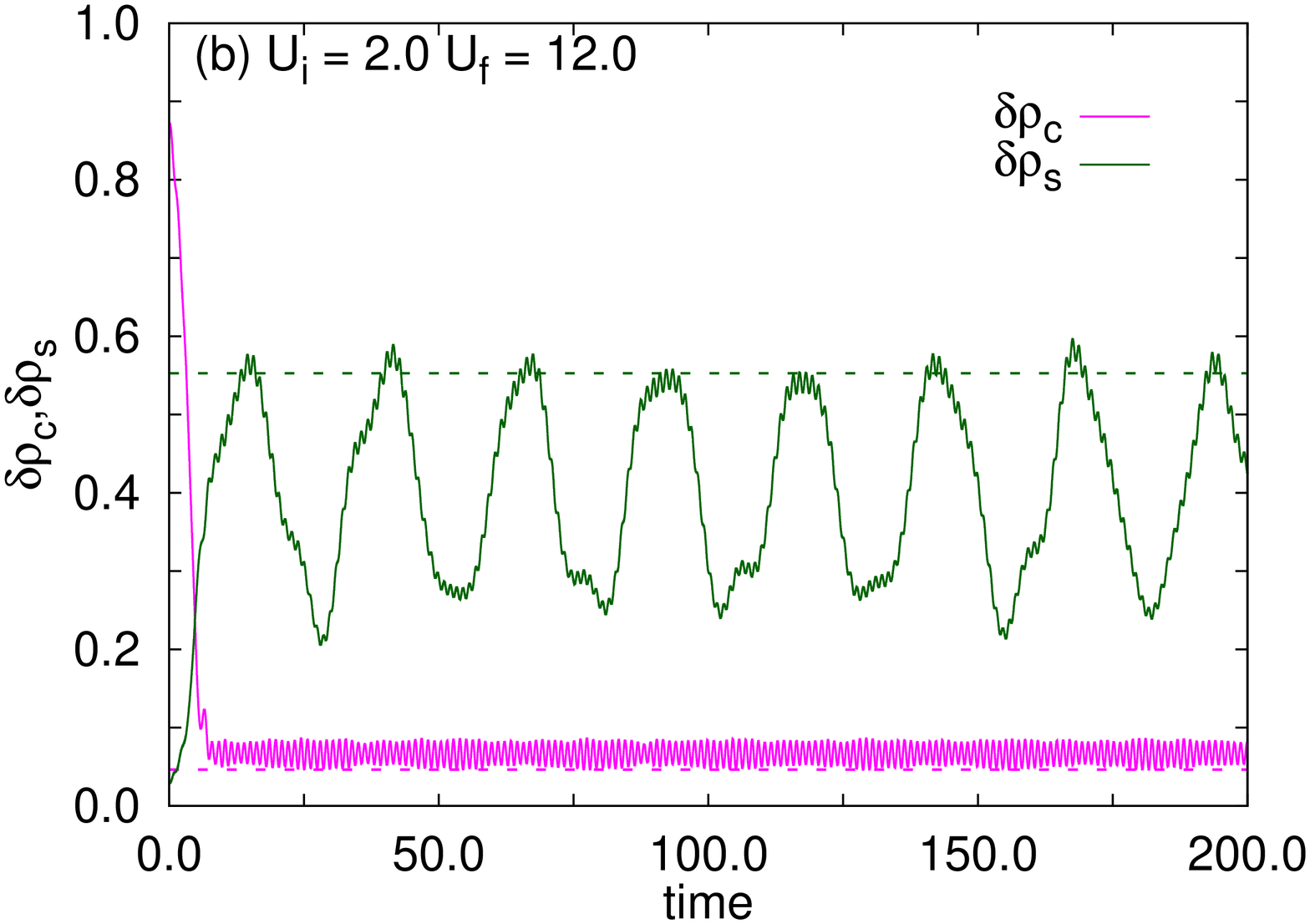}
\includegraphics[angle=-90,width=0.235\textwidth]{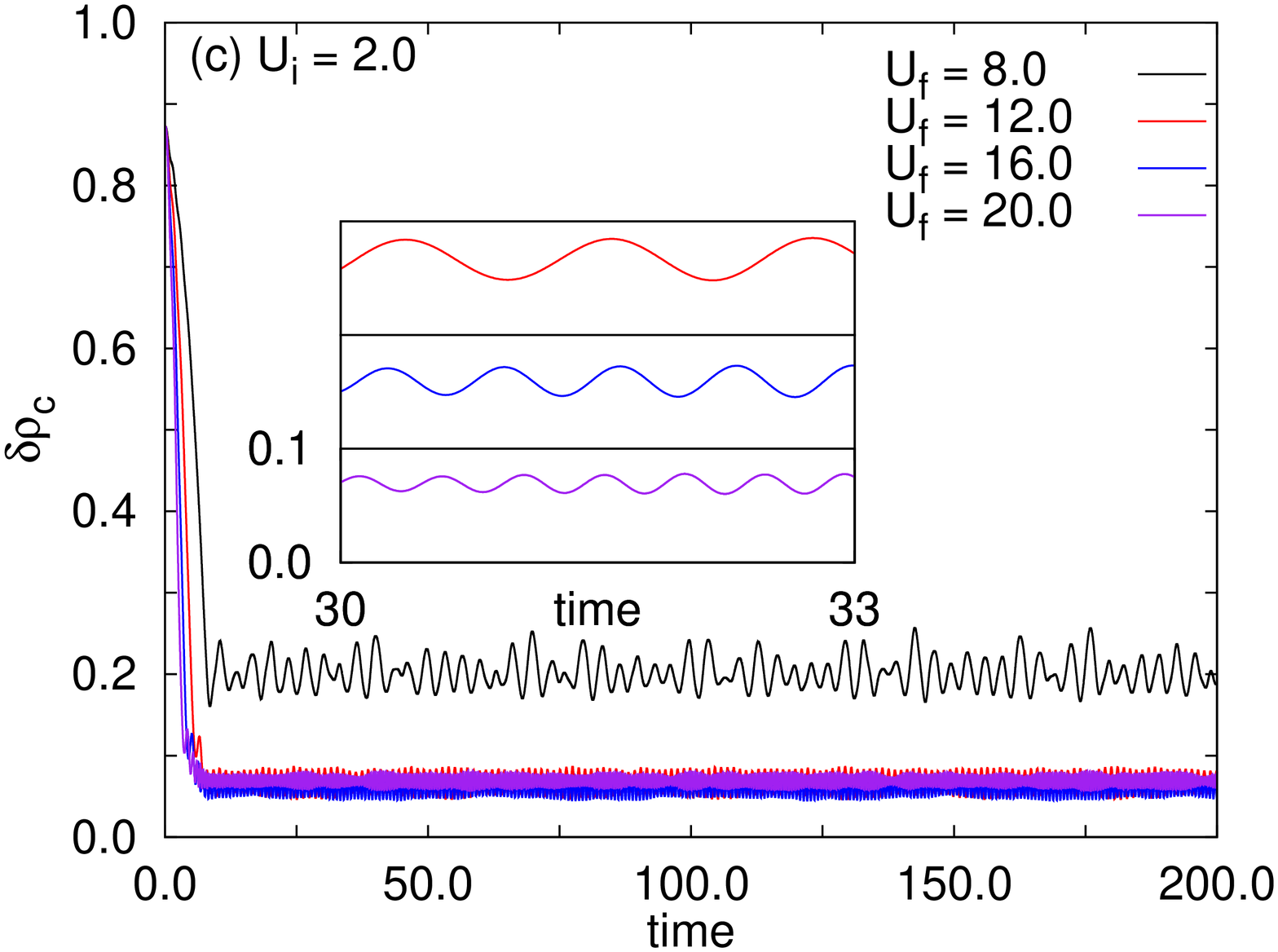}
\includegraphics[angle=-90,width=0.235\textwidth]{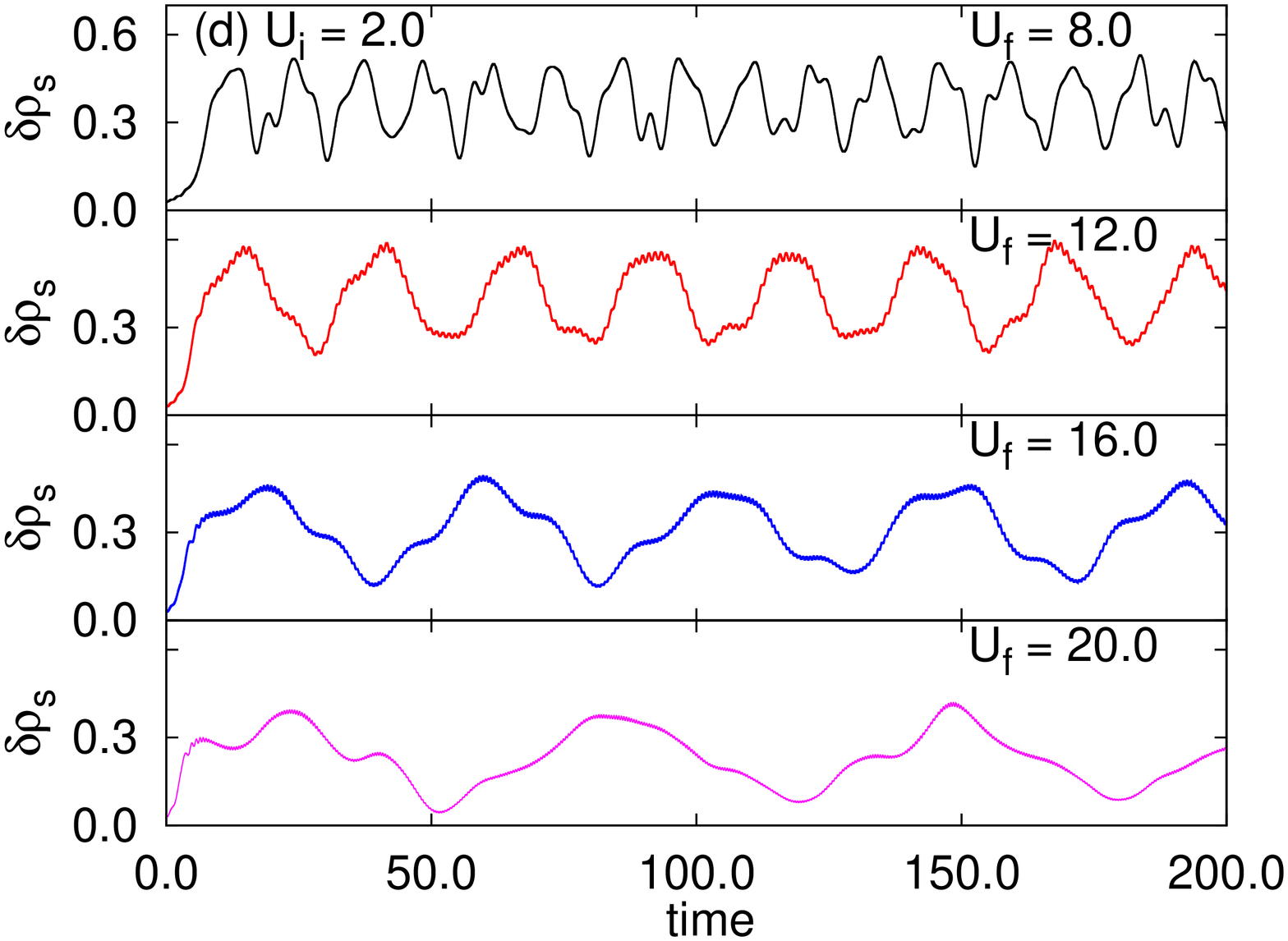}
\caption{
Time evolution of $\delta\rho_s$ and $\delta\rho_c$ for the quench protocol (a) $U_i=2.0 \rightarrow U_f=8.0$ and (b) $U_i=2.0 \rightarrow U_f=12.0$, respectively, where the quench time $t_q=8.0$. The dashed line represents the equilibrium $\delta\rho_c$ and $\delta\rho_s$ at zero temperature with $U=U_f$. (c) $U_i=2.0 \rightarrow U_f=8.0,\;12.0,\;16.0,\;20.0$, and the periods of the charge density order parameter are $3.27,\; 1.15,\; 0.69,\; 0.50$. The inset shows the oscillation behavior at $30<t<33$ for $U_f=12,\; 16,\; 20$. (d) $U_i=2.0 \rightarrow U_f=8.0,\;12.0,\; 16.0,\; 20.0$, and the periods of the spin density order parameter are $12.40,\; 24.32,\; 43.72,\; 63.53$.}
\label{fig4}
\end{figure*}

\subsection{Non-equilibrium quench dynamics - the strong Coulomb interaction limit}
In order to observe the oscillating behavior clearly, we set the final Coulomb interaction deep in the strong interaction regime. In Fig.\ref{fig4}(a), we plot the charge and spin order parameter for the Coulomb interaction quench from $U_i=2.0, U_f=8.0$. In the region $t > t_q$, the charge order parameter is oscillating around its equilibrium value $\delta\rho_c = 0.202$, while the spin order parameter is oscillating around $\delta\rho_s = 0.330$, which deviate from its equilibrium value at zero temperature $\delta\rho_s^{eq} = 0.438$. The effective temperature at $t \geq t_q$ is $T_{\mathrm{eff}} = 0.215$, where the order parameters are $\delta\rho_{c}^{eq} =  0.201$ and $\delta\rho_{c}^{eq} = 0.385$.
Upon further increasing the final Coulomb interaction $U_f=12.0$ while keeping the initial $U_i = 2.0$, leads to the results in Fig.\ref{fig4}(b). By inspecting the spin order parameter, an apparent oscillation period of $\delta\rho_{c}^{eq} =  0.350$ is observed. The charge order parameter as a function of time oscillates with a period $T=24.8$.

To illustrate the behavior of the charge and spin order parameters in the strong Coulomb interaction regime following a quench, we plot the spin-order parameter for the Coulomb interactions $U_f = 8.0,\; 12.0,\; 16.0,\; 20.0$ while fixing the initial Coulomb interaction as $U_i=2.0$ in Fig.\ref{fig4}(c-d).
Apparently, independent of the initial Coulomb interaction and the quenching time in protocols, the oscillation frequency of the charge order parameter increases monotonically with Coulomb interaction $U_f$.

The oscillation amplitude for the spin order parameter is clearly larger than the charge order parameter. This can be understood in the strong Coulomb interaction limit, where the effective Hamiltonian can be approximately described as an anisotropic XXZ spin model. In the effective spin Hamiltonian, the behavior is dominated by the spin excitations with energy scale $\mathcal{J}_{\mathrm{ex}}$ defined in Eq.\eqref{Jexdef}, which is in agreement with previous experiments\cite{Trotzky:sci2008}.

To confirm the behavior above, we further calculate the evolution of the two parameters for $U_i=2.0$ and $U_f=12.0$. The effective Hamiltonian, Eq.\eqref{Jexdef}, is an XXZ model with exchange coupling $\mathcal{J}_{\mathrm{ex}}$.
For the final Coulomb interaction $U_f = 12.0,\; 14.0,\; 16.0,\; 20.0$, the corresponding exchange interaction are $\mathcal{J}_{\mathrm{ex}} = 0.333,\; 0.263,\; 0.218,\; 0.165$, respectively. The oscillation periods of the charge density order parameter are $3.27,\; 1.15,\; 0.69,\; 0.50$ for $U_f=8.0,\; 12.0,\; 16.0,\; 20.0$. The periods of the spin density order parameter are $12.40,\; 24.32,\; 43.72,\; 63.53$, which increase monotonically with Coulomb interaction.

To understand the origin of the behavior with increasing $U_f$, one can can consider a two-site Hubbard model at half-filling in the strong Coulomb interaction limit, where the hopping terms are considered as a perturbation.
The energy levels from lowest to highest are the singly occupied singlet and triplet, and the other two energy levels have a double occupancy on one of the sites.
The oscillations of the spin order parameter will be determined mainly by the excitation energy between the singlet and the triplet, which is the effective Hund's coupling energy, $\mathcal{J}_{\mathrm{ex}} \sim U^{-1}$.
By contrast, the oscillations of the charge order will incorporate the excitations between singly occupied and doubly occupied states, where the energy difference will be mainly determined by Coulomb interaction $U$.  This explains the corresponding monotonically  decreasing and increasing oscillation period trends of the charge and spin order parameters with increasing $U_f$.

\section{Conclusion}
\label{concl}
In this paper, the exact diagonalization (time dependent Lanczos) method is used to study the quench dynamics in the one-dimensional ionized Hubbard model with mass imbalance.
To check the validity of ED used in studying the system, we first studied the phase diagram in equilibrium and compared it with the phase diagram calculated using DMRG or Hartree-Fock mean-field methods.
Qualitatively, the ED calculation shows that the transition between the band insulator and correlated insulator is of second order, which is consistent with DMRG calculation.

The phase diagram in the Coulomb interaction and crystal field plane $U-\Delta$ is studied. The critical Coulomb interaction deviate from the DMRG result for a small $\Delta<0.25$ region.
The phase diagram in the $U-\eta$ plane ($\eta$ is mass imbalance) is studied. In comparison with DMRG, we find ED works well for relative large $\eta\leq 0.75$.
Furthermore, finite size effects are studied by considering 10-site and 14-site chains. We find that increasing the number of sites will greatly improve the agreement with DMRG. We choose the mass imbalance and the crystal field parameters as $\eta = 0.75$ and $\Delta = 3.0$ in this paper, respectively, where the phase transition point from band insulator to correlated insulator is $U_c=7.1$.

Focusing on the non-equilibrium evolution after a Coulomb quench, we study the dependence on the quenching time $t_q$ for a fixed initial and final Coulomb interaction $U_i=6.0$ and $U_f=8.0$, where $U_i=6.0$ is in the band insulating regime in equilibrium and $U_f$ is situated in the correlated insulating regime. By inspecting the time evolution of the charge and spin order parameters, we observe that the two order parameters exhibit different oscillation behaviors, which depend on the quenching time. In general, a monotonically increasing (decreasing) spin (charge) order parameter at short times followed by an approximate oscillating behavior at long times is observed.

In the long time regime, the order parameters  oscillate around their thermalized equilibrium value, where an effective temperature is defined. Furthermore, the effective temperature will decrease monotonically with quenching time for fixed initial and final Coulomb interaction $U_i$ and $U_f$, where an approximate adiabatic evolution is observed for very large $t_q$.

Finally, we study the parameter region where $U_i = 2.0$ and $U_f$ deep in the correlated insulating region. The spin and charge order parameters will oscillate with time in the long time regime.
The oscillating frequency of charge order parameter will increase monotonically with the Coulomb interaction.
In contrast, the spin order parameter will decrease monotonically with the Coulomb interaction. The oscillating frequency is independent with quenching time in the protocol and the initial Coulomb interaction.

In summary, we study the non-equilibrium evolution of the mass imbalanced ionic Hubbard model driven by a Coulomb interaction quench. Our results show that the dynamical evolution of physical observable exhibit different behaviors depending on the quench protocol, where the effective temperature decreases with increasing quench time. When the final Coulomb interaction strength is situate deep in the correlated regime, the oscillation period of the spin (charge) order parameter will increase (decrease) monotonically with Coulomb interaction strength, which is independent of quench protocol. Our results can be tested experimentally in cold atom optical lattices, and may offer strategies to engineer the relaxation behavior of interacting quantum many-particle systems.

\acknowledgements
We acknowledge helpful discussions with Xiaojun Zheng, Huan Li and Yun Guo.
We gratefully acknowledge funding from the National Natural Science Foundation of China (Grant No. 11904143, 12174168, 12047501, 11764017, 11547184),
the Natural Science Foundation of Guangxi Province Grant No. 2020GXNSFAA297083, GuiKe AD20297045.
G.A.F. gratefully acknowledges funding from US National Science Foundation grant DMR-2114825.

\appendix
\setcounter{equation}{0}
\renewcommand\theequation{A\arabic{equation}}
\setcounter{figure}{0}
\renewcommand\thefigure{A\arabic{figure}}
\begin{figure*}[ht]
\centering
\includegraphics[angle=-90,width=0.33\textwidth]{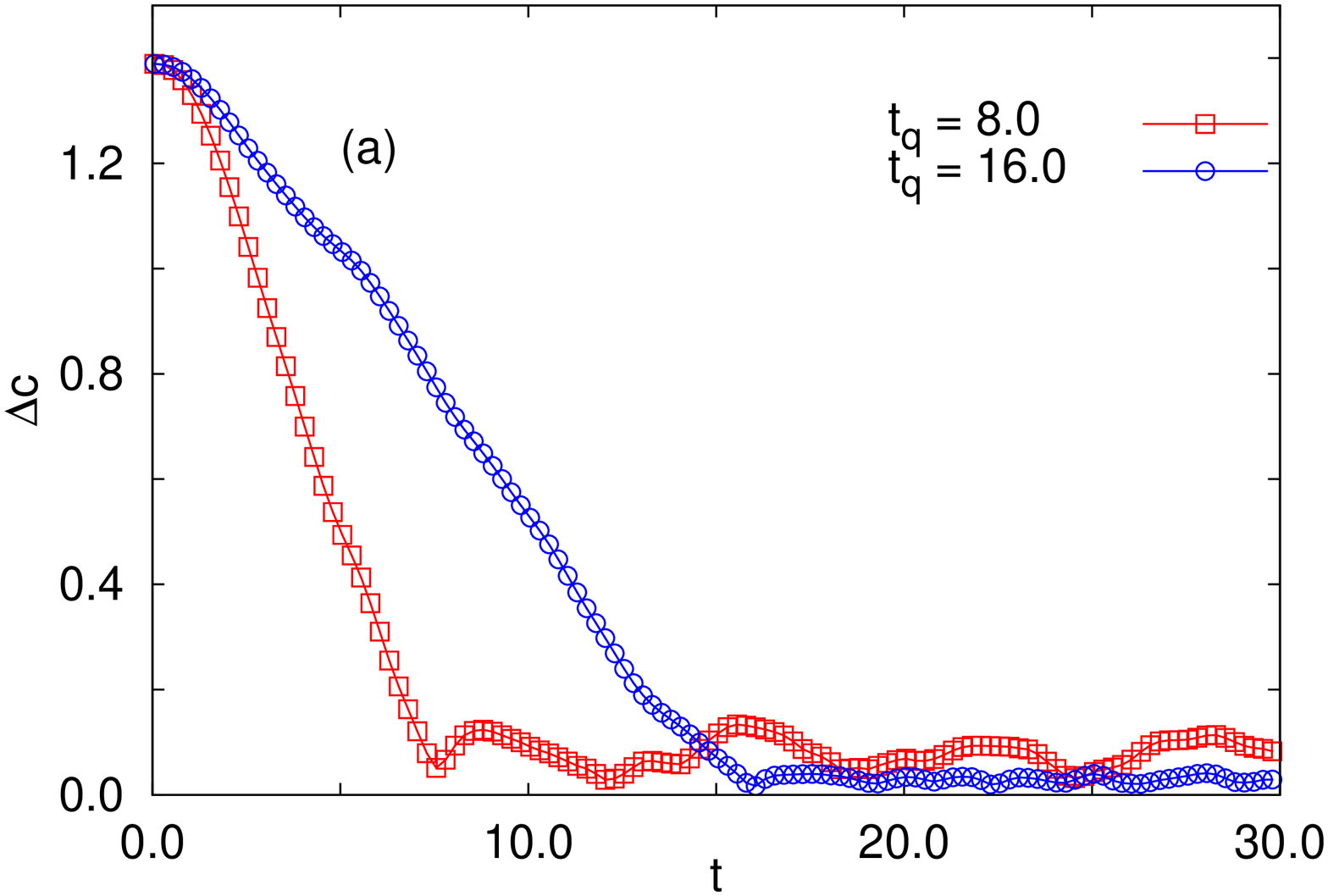}
\includegraphics[angle=-90,width=0.33\textwidth]{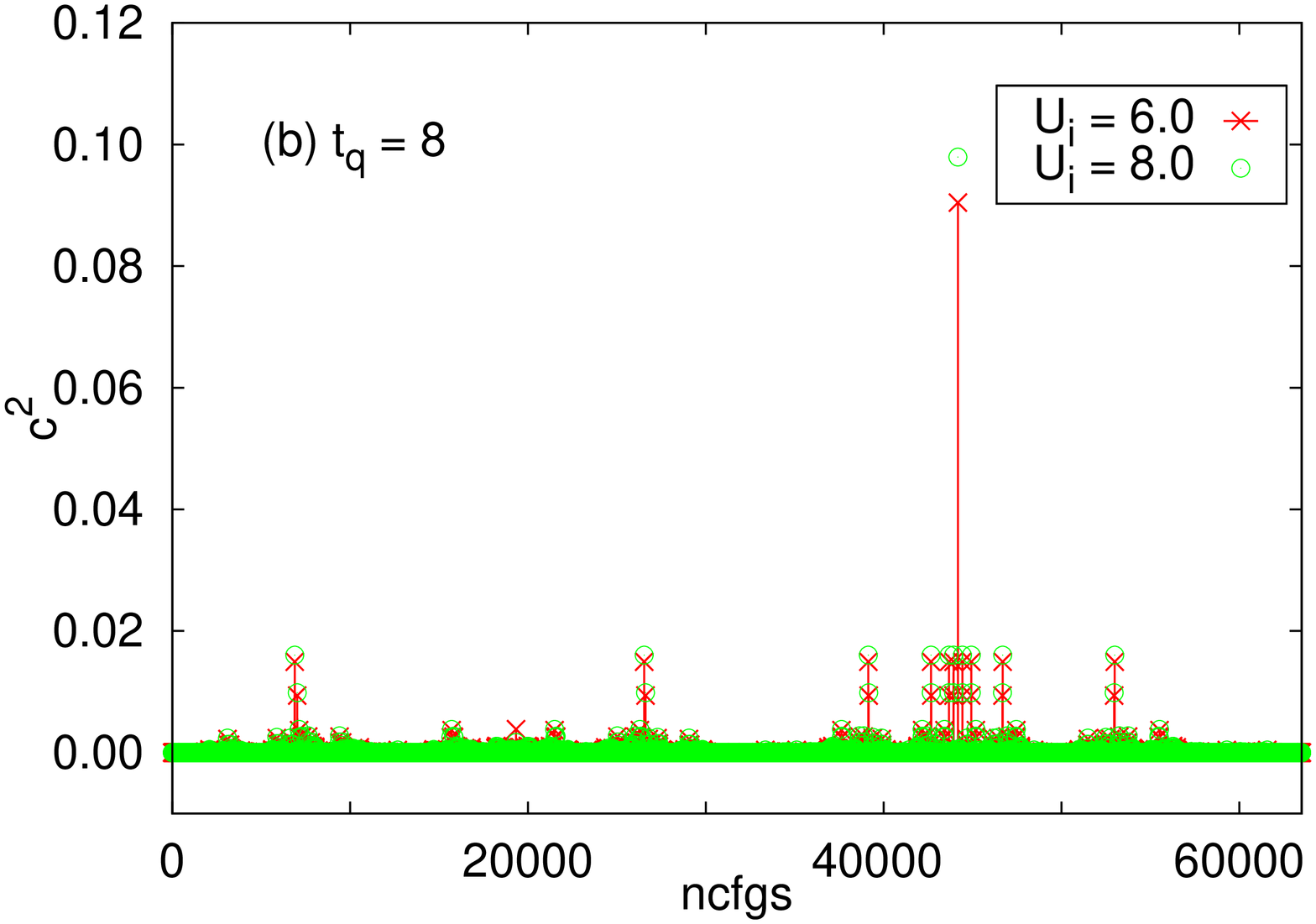}
\includegraphics[angle=-90,width=0.33\textwidth]{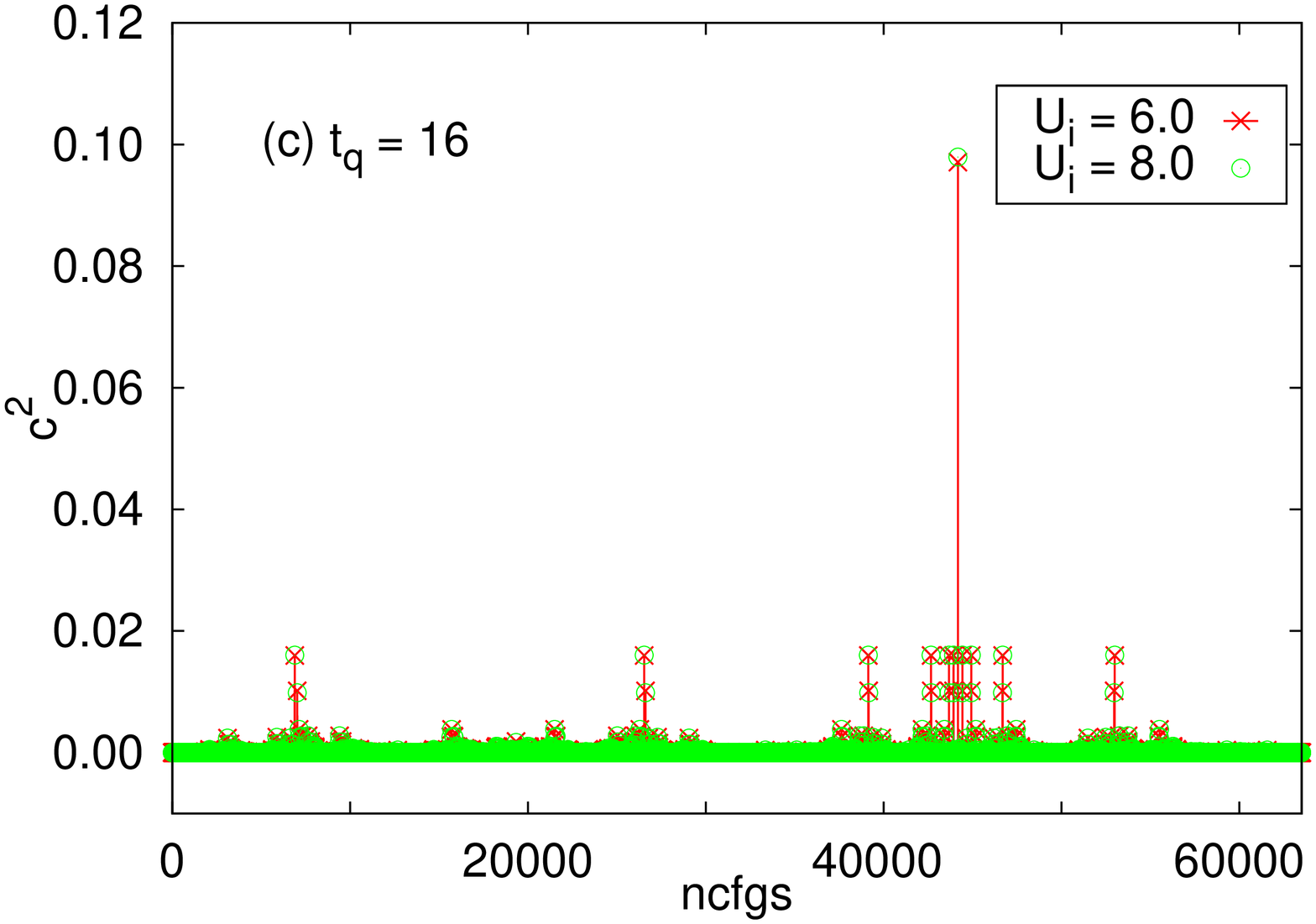}
\caption{Set $U_i=6.0$ and $U_f=8.0$. (a) The time evolution of $\Delta c$ are plotted as a function of time with the quenching time $t_q=8.0$ and $t_q=16.0$, respectively.
(b)-(c) The probability distribution of each Fock basis in many body Hilbert space for $|\langle I|\Psi_g^f\rangle|^2$ and $|\langle I |\Psi(t)\rangle|^2$ defined in Eq.\eqref{eq:prob}
are plotted with quenching time $t_q = 8$ (b) and $t_q = 16$ (c), respectively.}
\label{fig5}
\end{figure*}
\begin{figure*}[ht]
\centering
\includegraphics[angle=-90,width=0.245\textwidth]{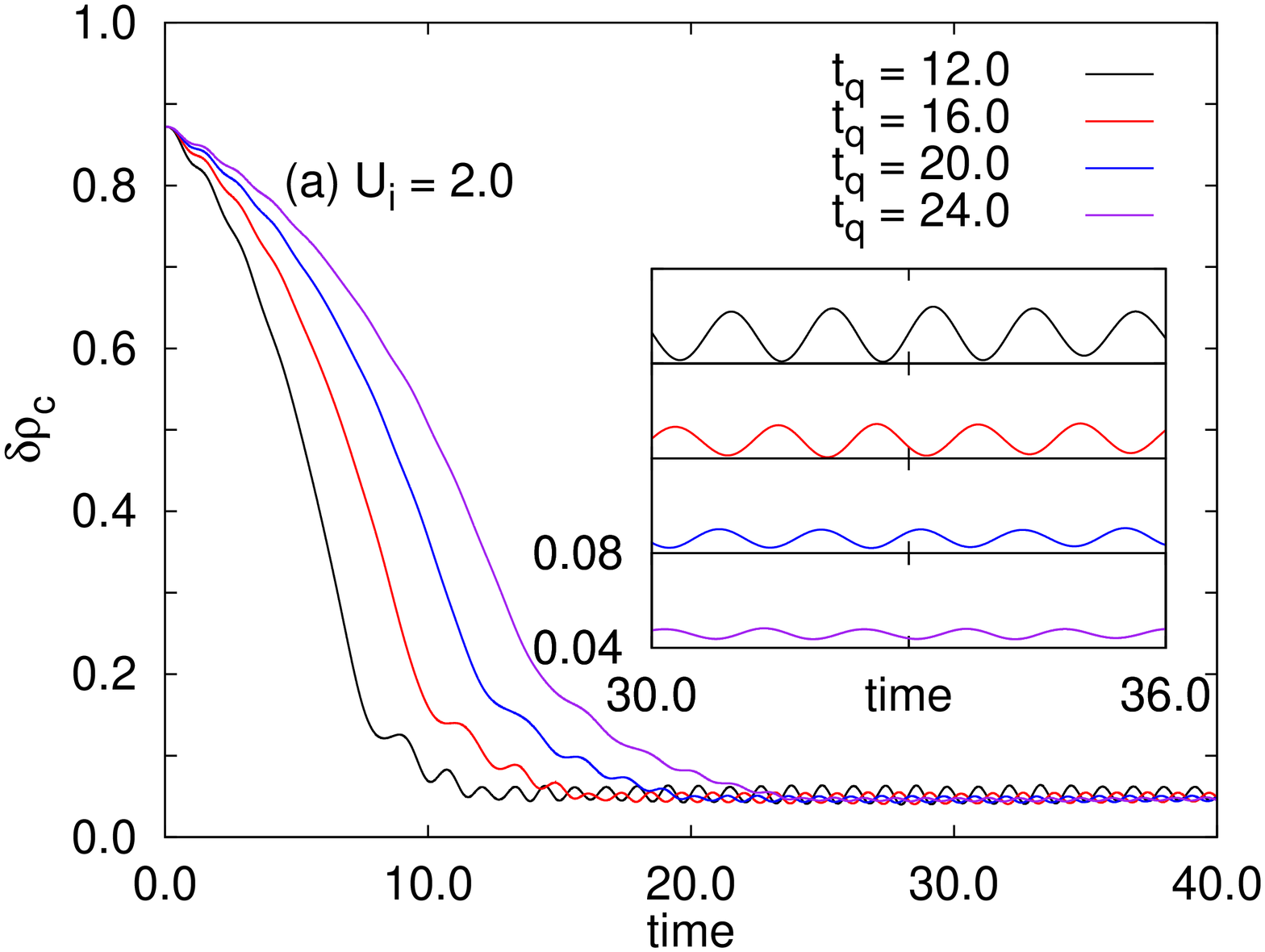}
\includegraphics[angle=-90,width=0.245\textwidth]{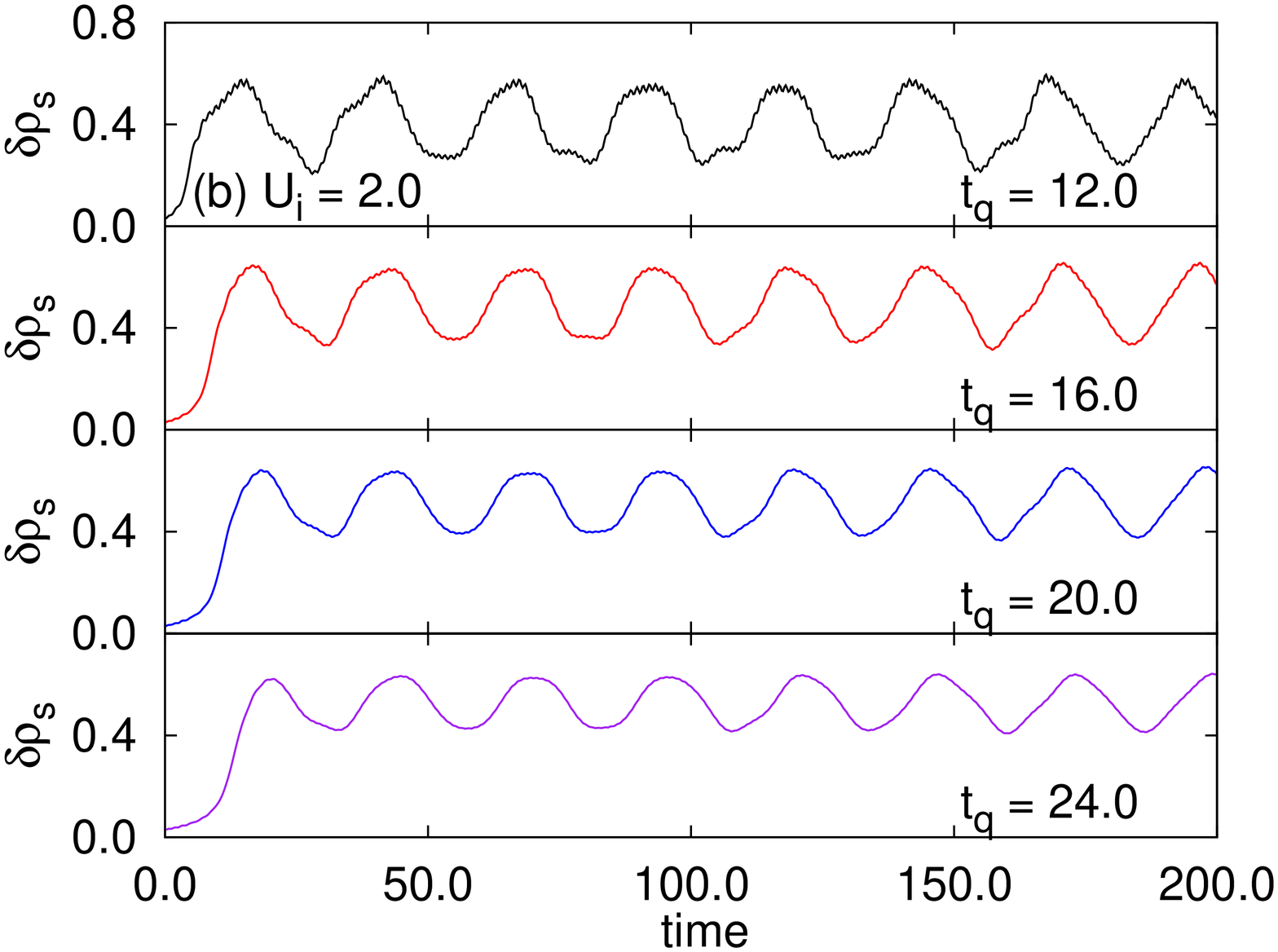}
\includegraphics[angle=-90,width=0.245\textwidth]{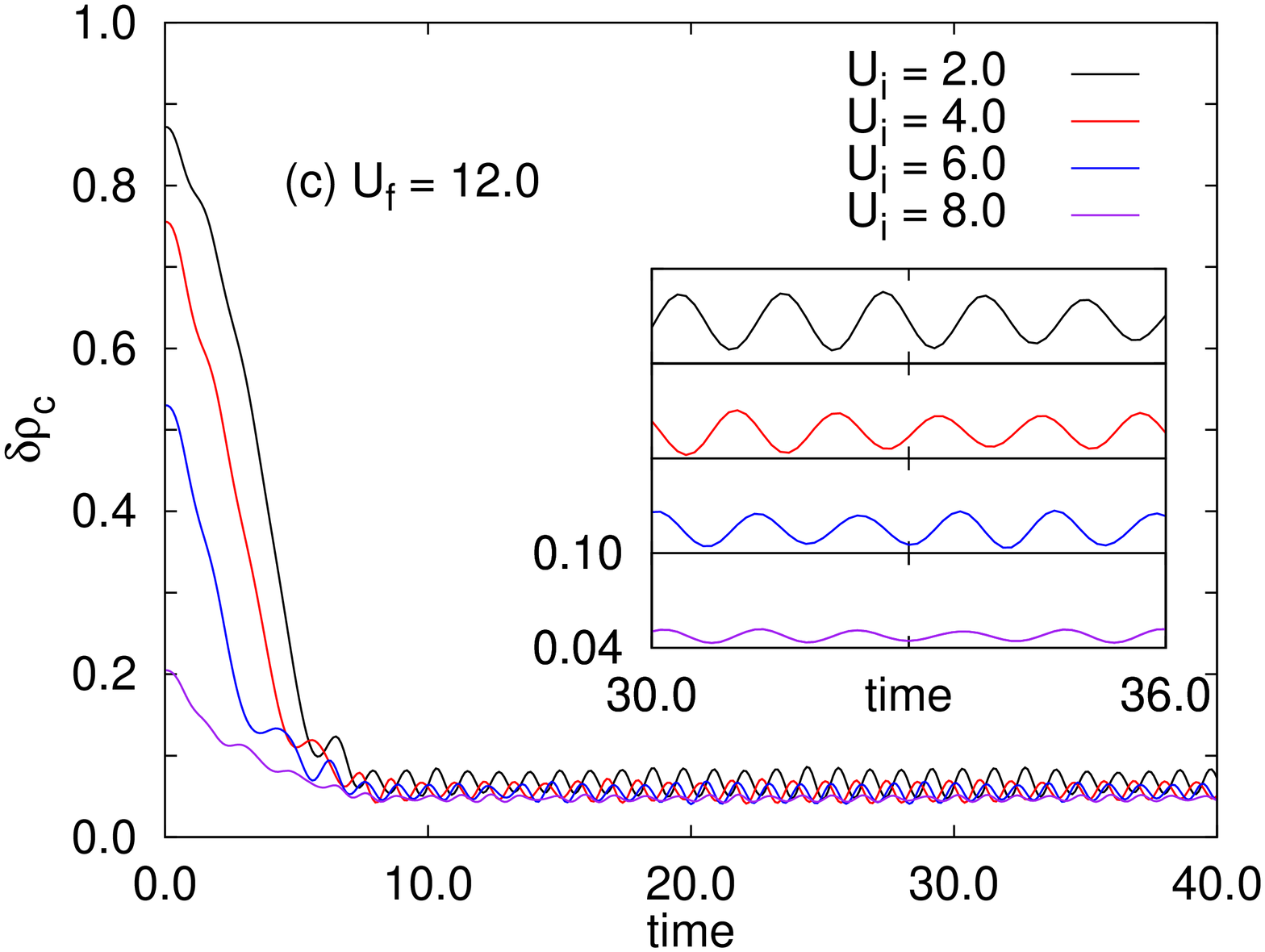}
\includegraphics[angle=-90,width=0.245\textwidth]{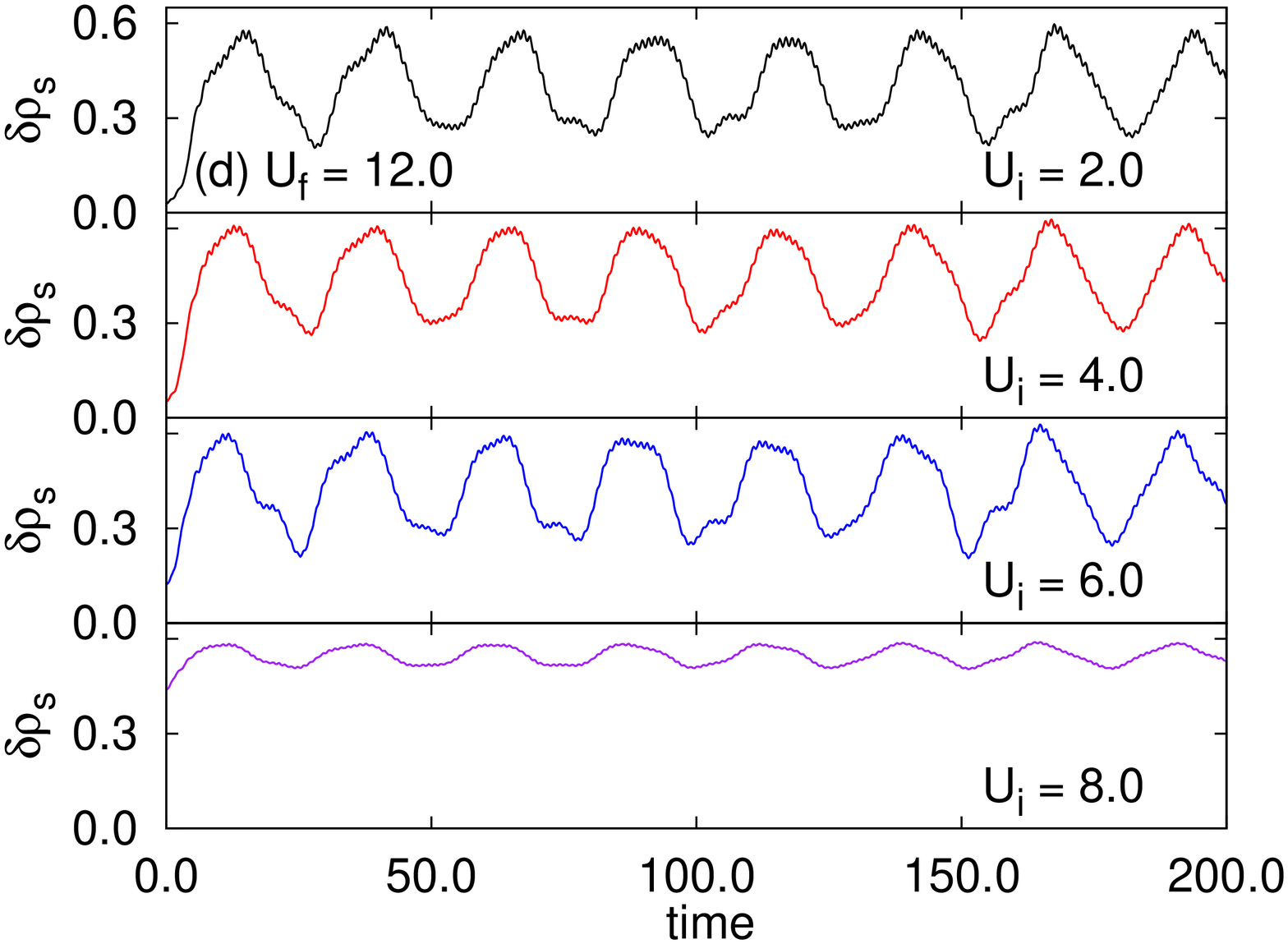}
\caption{(a-b)
Time evolution of the charge and spin order parameter $\delta\rho_c$ and $\delta\rho_s$ after the Coulomb interaction quench
where the initial and final Coulomb interactions are set as $U_i$ = 2.0 and $U_f$ = 12.0, respectively.
The quench protocols with different quenching time $t_q=12.0,\; 16.0,\; 20.0,\; 24.0$ are applied, respectively.
(c-d) Time evolution of $\delta\rho_c$ and $\delta\rho_s$ for different quench protocols with fixed final Coulomb interaction $U_f=12.0$ and quenching time $t_q=8$ while the initial Coulomb interaction is changed with $U_i = 2.0,\;4.0,\;6.0,\;8.0$.}
\label{fig6}
\end{figure*}

\section{The effect of long quenching time on the system state}
The time evolution of the ground state wave function that approximates adiabatic evolution under a long quenching time is studied here.
In Fig.\ref{fig2}(e-f), the quenching time in protocols are $t_q=8.0$ and $t_q=16.0$, respectively.
The effective temperature of the systems are $T_\mathrm{eff} = 0.105, 0.008$ (close to 0 K), where the non-equilibrium ramp process can possibly be approximated as adiabatic process.

To characterize the difference between the non-equilibrium evolution studied in our work and an adiabatic process, we define a parameter $\Delta c$ to measure the difference,
\begin{align}
  \Delta c(t) = \sum_I \left|\langle \Psi(t)|I\rangle\langle I |\Psi(t)\rangle - \langle \Psi_g^f|I\rangle\langle I| \Psi_g^f\rangle\right|,
  \label{eq:prob}
\end{align}
where $|\Psi_g^f\rangle$ is the equilibrium ground state wave function of the final Hamiltonian with $U=U_f$, $|\Psi(t)\rangle$ corresponds to non-equilibrium wave function at time $t$, and $I$ labels the $i$-th Fock basis state in the many-body Hilbert space.

In Fig.\ref{fig5}(a), the parameter $\Delta c$ with quenching times $t_q = 8.0, 16.0$ are plotted as a function of time.
It is observed that the parameter $\Delta c$ decreases monotonically and approaches 0. To further understand the detailed difference between the two wave functions, we plot the probabilities of each Fock basis state in Fig.\ref{fig5}(b-c) for the equilibrium wave function $|\Psi_g^f\rangle$ and the non-equilibrium wave function $|\Psi(t)\rangle$ at $t=30$ with different quenching times $t_q=8.0$ and $t_q=16.0$, respectively.
The basis with highest probability is $|\downarrow,\uparrow,\downarrow,\uparrow,\cdots ,\downarrow,\uparrow\rangle$.

\renewcommand\theequation{B\arabic{equation}}
\setcounter{figure}{0}
\renewcommand\thefigure{B\arabic{figure}}

\section{Effects of quenching time and initial Coulomb interaction on the oscillation period of the order parameter}
Here we study the effect of quenching time $t_q$ on the oscillation period of the order parameter of the system under strong interactions.
By setting the initial and final Coulomb interaction as $U_i=2.0$ and $U_f=12.0$, we plot the time evolution of the charge and spin order parameters with different quenching times $t_q = 12.0,\; 16.0,\; 20.0,\; 24.0$ in Fig.\ref{fig6}(a-b).
By solving Eq.\eqref{Teff}, the effective temperatures are $T_{\mathrm{eff}} = 0.345,\; 0.114,\; 0.089,\; 0.020$, respectively.
The oscillation period of the charge density and spin order parameters are $T_{\mathrm{CDW}} \approx 1.16$ and $T_{\mathrm{SDW}} \approx 26.02$, which are independent of the quenching time.

In addition, we study the oscillation behavior of the charge and spin order parameters by fixing the final Coulomb interaction $U_f=12$ in the strong Coulomb interaction regime with quenching time $t_q =8$ while changing the
initial Coulomb interaction $U_i=2.0,\; 4.0,\; 6.0,\; 8.0$. The two order parameters as a function of time are plotted in Fig.\ref{fig6}(c-d).
Our numerical results show that the oscillation periods of the charge density and spin density order parameters are about $T_{\mathrm{CDW}}\approx1.15$ and $T_ {\mathrm{SDW}} \approx 24.03$, which are independent of the initial Coulomb interaction strength.

\bibliography{massimbalance}
\end{document}